\documentclass[aps,prd,nofootinbib,superscriptaddress]{revtex4} 
\usepackage{graphics}
\usepackage{epsfig}
\usepackage[english]{babel}
\usepackage{amsmath} 
\usepackage{verbatim} 
\usepackage{mathrsfs}
\usepackage{amsfonts}
\usepackage{amssymb}
\usepackage[latin1]{inputenc}
\usepackage{hyperref}

\newcommand{\tc}{\eta_{\vec{k}}^c}
\newcommand{\beq}{\begin{equation}}
\newcommand{\eeq}{\end{equation}}
\newcommand{\nk}{\vec{k}}
\newcommand{\dphi}{\delta \phi}
\newcommand{\x}{\vec{x}}

\newcommand{\bra}{\langle}
\newcommand{\ket}{\rangle}

\newcommand{\mH}{\mathcal{H}}

\newcommand{\barr}{\begin{eqnarray}}
\newcommand{\earr}{\end{eqnarray}}
\newcommand{\bea}{\begin{eqnarray*}}
\newcommand{\eea}{\end{eqnarray*}}

\newcommand{\half}{\frac{1}{2}}

\newcommand{\cre}{\hat{a}^{\dagger}}    

\newcommand{\ann}{\hat{a}}

\begin{document}

\title{Quantum collapse as source of the seeds of cosmic structure during the radiation era}

\author{Gabriel Le\'{o}n}
\affiliation{Departamento de F\'{\i}sica, Facultad de Ciencias Exactas y Naturales, Universidad de Buenos Aires, Ciudad Universitaria - PabI, Buenos Aires 1428, Argentina}

\author{Susana J. Landau}
\affiliation{Departamento de F\'{\i}sica, Facultad de Ciencias Exactas y Naturales, Universidad de Buenos Aires and IFIBA, CONICET, Ciudad Universitaria - PabI, Buenos Aires 1428, Argentina\thanks{Member of  the Carrera del 
                               Investigador Cient\'{\i}fico y Tecnol\'ogico, 
                               CONICET}}

\author{Mar\'{\i}a P\'{\i}a Piccirilli}
\affiliation{Grupo de Astrof\'{\i}sica, Relatividad y Cosmolog\'{\i}a, Facultad de Ciencias Astron\'{o}micas y Geof\'{\i}sicas, Universidad Nacional de La Plata, Paseo del Bosque S/N 1900 La Plata, Pcia de Buenos Aires, Argentina}

\begin{abstract}
The emergence of the seeds of cosmic structure, from a perfect isotropic and homogeneous Universe, has not been clearly explained by the standard version of inflationary  models as the dynamics involved preserve the homogeneity and isotropy at all times. A proposal that attempts to deal with this problem, by introducing ``the self-induced collapse hypothesis,'' has been introduced by D. Sudarsky and collaborators in previous papers. In all these works, the collapse of the wave function of the inflaton mode is restricted to occur during the inflationary period. In this paper, we analyze the possibility that the collapse happens during the radiation era. A viable model can be constructed under the condition that the inflaton field variable must be affected by the collapse while the momentum variable can or cannot be affected.  Another condition to be fulfilled is that the time of collapse must be independent of $k$. However, when comparing with recent observational data, the predictions of the model cannot be distinguished from the ones provided by the standard inflationary scenario.  The main reason for this arises from the requirement that primordial power spectrum obtained for the radiation era matches the amplitude of scalar fluctuations consistent with the latest CMB observations. This latter constraint results in a limit on the possible times of collapse and  ensures  that the contribution of the inflaton field to the energy-momentum tensor is negligible compared to the contribution of the radiation fields.  
\end{abstract}


\maketitle

\section{Introduction}
\label{intro}

Observations of the Cosmic Microwave Background (CMB) radiation are one of the most powerful tools to study the early Universe, also they can provide precise measurement of the cosmological parameters. Starting with COBE's groundbreaking detection, in the past two decades there has been a major improvement in the measurement of microwave background temperature fluctuation. On the other hand, recent observations of the CMB power spectrum, e.g. the release of \emph{Planck} data \cite{Planckcls13} and  the recent claim about the detection of B-modes originated by primordial gravitational waves \cite{BICEP2}, has strengthened the theoretical status of inflationary scenarios among cosmologists.

In the standard (and the simplest) inflationary scenario, the origin of structures in our Universe like galaxies and clusters of galaxies is explained by assuming a stage described by an accelerating (nearly de Sitter) expansion driven by the potential of a single scalar field, and from its quantum fluctuations characterized by a simple vacuum state.  In particular, the quantum fluctuations transform into the classical statistical fluctuations that represent the seeds of the current cosmic structure. However, the usual account for the origin of cosmic structure is not  fully  satisfactory  as it   lacks  a physical mechanism capable of   generating the inhomogeneity and anisotropy of our Universe,  from an exactly homogeneous and isotropic initial state associated with the early inflationary regime. This issue has been analyzed in previous papers \cite{PSS06,Shortcomings,LLS13} and one key aspect of the problem is that there is no satisfactory solution within the standard physical paradigms of quantum unitary evolution because this kind of  dynamics is not capable to  break the initial symmetries of the system. To handle this shortcoming, a proposal has been developed by D. Sudarsky and collaborators \cite{PSS06,Sudarsky07,US08,Leon10,Leon11,DT11,LSS12,CPS13,LLS13}. In this scheme, a new ingredient is introduced into the inflationary scenario: \emph{the self-induced collapse hypothesis}. The main assumption is that,  at a certain stage in the cosmic evolution, there is an induced jump from the original quantum  state characterizing the particular mode of the quantum field; after the jump, the quantum state is inhomogeneous and anisotropic or more precisely it must not be an eigen-state of the linear and angular momentum operators. This process  is similar to the quantum mechanical reduction of the wave function associated with a measurement. However, in our scheme,  there is no external measuring device or observer  (as there is nothing in the situation we are considering that could be called upon to play such a role). The hypothesis concerning an observer independent collapse of the wave function has been proposed and analyzed in the community working on quantum foundations:  The continuous spontaneous localization (CSL)  model \cite{pearle1989}, representing  a   continuous  version of the Ghirardi-Rimini-Weber  model \cite{ghirardi1985}, and the proposals of Penrose \cite{penrose1996} and Di\'osi \cite{diosi1987,diosi1989} addressing gravity as the main agent for triggering the reduction of the wave function, are among the main schemes attempting to model the physical mechanism of a self-induced collapse (for more recent examples  see Refs. \cite{weinberg2011,bassi2003}).


Therefore, by considering a self-induced collapse (in each mode) of the inflaton wave function, the inhomogeneities and anisotropies arise at each particular length scale. As a consequence of this modification of the inflationary scenario, the predicted primordial power spectrum is modified and also the CMB fluctuation spectrum.  Previous works \cite{PSS06,Shortcomings,LLS13,DT11} have extensively discussed both the conceptual and formal aspects of this  new proposal, and we refer the reader to the references. However, we would like to comment on an important point, namely the characteristics of the state into which such jump occurs. As mentioned previously, the quantum state must not be an homogeneous and isotropic state. One could then assume a particular collapse mechanism, which would lead to such post-collapse state, and then calculate the corresponding observables in that state. The question now would be: which are the appropriate observables for the problem at hand that emerge from the quantum theory? 

One possible approach would be to assume that both--metric and matter--perturbations are well characterized by a quantum field theory constructed on a classical unperturbed background; in the context of inflation, this approach corresponds to the quantization of the so-called Mukhanov-Sasaki variable, which then is used to yield predictions for the observational quantities (e.g. the spectrum of the temperature anisotropies). Therefore, if one assumes a particular collapse mechanism, which  somehow modifies the standard unitary evolution given by Schroedinger's equation, then the dynamic of the observables, in terms of the Mukhanov-Sasaki variable, would be modified directly; this scheme was developed in Refs. \cite{jmartin,tpsingh} for the inflationary Universe.

Another possible approach to relate the quantum degrees of freedom with the observational quantities, is to rely on the semiclassical gravity picture; within this framework, the metric perturbations are always described in a classical way, while the matter degrees of freedom are modeled by a quantum field theory in a curved classical background. Then, by using Einstein's semiclassical equations $G_{ab} = 8 \pi G  \bra \hat{T}_{ab} \ket$, one relates the quantum matter perturbations with the corresponding ones from the classical metric. Nevertheless, assuming a particular collapse mechanism, which once again can be thought as a modification of standard Schroedinger's equation, would not affect the dynamics of the metric perturbation; indeed, the dynamics of the modes characterizing the quantum field would be modified,  but since the metric perturbation is always a classical object, its dynamics is not given by the modified  Schroedinger's equation. Assuming a particular collapse mechanism, would only modify the initial conditions of the motion equation for the metric perturbation, which again is always described at the classical level; in the context of inflation, this was analyzed in Ref. \cite{CPS13}.

In this work, we will take the semiclassical gravity approach, since (as will be argued in the paper) it presents a clear picture of how the inhomogeneities and anisotropies are born from the quantum collapse. Moreover, since the consideration of a particular collapse mechanism will not alter the dynamics of the classical quantities, we can characterize the post-collapse state in a generic way. In particular, we will follow the pragmatical approach first proposed in \cite{PSS06} in which one describes the collapse by characterizing the expectation values of the quantum field variable and its momentum in the post-collapse state.  In Refs. \cite{PSS06,US08,LSS12} two schemes were considered; one in which, after the collapse, both expectation values are randomly distributed within  their respective ranges of uncertainties in the pre-collapsed state, and another one in which it is only the conjugate momentum that changes its expectation value from zero to a value in its corresponding range as a result of the collapse. In this paper, we will also consider the possibility that only the field variable changes its expectation value after the collapse.

On the other hand, in all previous works \cite{PSS06,US08,DT11,CPS13} the self-induced collapse of the inflaton wave function is restricted to happen at the inflationary stage of the Universe. However, there is no  reason for this restriction, apart from the observational limits imposed by the CMB data.   As matter of fact, the idea of generating the primordial curvature perturbation after the inflationary era has ended is not a new proposal; earlier works based on the curvaton scenario deal with such picture \cite{curvaton1,curvaton2,curvaton3}, and  even in recent works \cite{curvaton4} the curvaton model is still, under certain assumptions,  considered as a viable option for generating the curvature perturbations. Moreover, in a model  by R.M. Wald \cite{wald},  the density perturbations can be achieved even if there was no inflationary regime at all. The aim of the present paper is to analyze the possibility that the primordial curvature perturbation can be generated by a self-induced collapse of the wave function of the inflaton field, but with the additional hypothesis that such collapse occurs during the radiation dominated epoch. We analyze three different possibilities for the post-collapse state of the wave function in a radiation dominated background. As we will show, it is possible to obtain a viable model, i.e. a nearly scale invariant power spectrum. Nevertheless, when comparing the model's prediction with recent data from the CMB temperature and temperature-polarization spectra, the predictions of the collapse model are essentially indistinguishable from the ones given by the traditional slow-roll inflationary scenario provided by a single scalar field.

The paper is organized as follows: In Sec. \ref{classical}, we present the action of the model and solve Einstein's semiclassical equations. In Sec. \ref{quantum}, we perform the quantization of the inflaton field in a radiation dominated background. In Sec. \ref{collapseschemes}, we introduce the collapse hypothesis for three different choices of the post-collapse state: i) the collapse affects only the field variable, ii) the collapse affects only the momentum variable, iii) the collapse affects both the field and momentum variable. In Sec. \ref{oquantities}, we relate the CMB observational quantities with the primordial spectrum modified with the collapse hypothesis. In Sec. \ref{analisis}, we analyze, from the theoretical point of view, the viability of the power spectrum obtained from each one of  three proposed collapse schemes. In Sec. \ref{camb}, we present an analysis where recent observational data is used to examine the validity of the predicted power spectrum. Finally, in Sec. \ref{discussion}, we end with a brief discussion of our conclusions.   Regarding notation and conventions, we will work with signature $(-,+,+,+)$ for the metric; primes over functions will denote derivatives with respect to the conformal time $\eta$, and we will use units where $c=\hbar=1$ but keep the gravitational constant $G$.

 \section{Classical analysis}
\label{classical}
 
The background space-time will be described by a spatially flat Friedmann-Robertson-Walker (FRW) radiation dominated Universe. The action of the theory is: 

\begin{equation}\label{action}
S = S_{\text{rad}} + S_G + S_{\text{inf}},  
\end{equation}
with $S_G$ is the standard action describing the gravity sector; $S_{\text{rad}}$ represents the action of the dominant type of matter, which in our case would be  radiation type of matter and $S_{\text{inf}}$ is the action of a single scalar field $\phi$  minimally coupled to gravity and with an appropriate potential representing the inflaton:

\beq
S_{\text{inf}} = \int d^4x \sqrt{-g} \bigg[  -\half \nabla_a \phi \nabla_b \phi g^{ab} - V[\phi] \bigg].
\eeq
Varying the action \eqref{action} with respect to the metric yields Einstein's equations

\beq
G_{ab} = 8 \pi G (T_{ab}^{\text{rad}} + T_{ab}^{\text{inf}}).
\eeq
The energy-momentum tensor for the inflaton can be written as:

\beq
T^{a \text{ inf}}_{b} = g^{ac} \nabla_c \phi \nabla_b \phi + \delta^a_b \left( \half g^{cd} \nabla_c \phi \nabla_d \phi -V[\phi] \right).
\eeq
Since we will work in a radiation dominated Universe, the contribution of $T_{ab}^{\text{inf}}$ to the total energy-momentum tensor  should be negligible, i.e. $T_{ab}^{\text{inf}} \ll T_{ab}^{\text{rad}}$. As usual, we separate the fields into a ``background'' part, taken to be homogeneous and isotropic, but in this case we have FRW radiation dominated Universe instead of quasi de-Sitter (inflaton) driven Universe, and the perturbations. In this way, the metric and the energy-momentum tensor field are  written as: $g = g_0 + \delta g$ and $T_{ab} = T_{ab}^{(0)} + \delta T_{ab}$. One can then apply perturbation theory to Einstein's equations. Nevertheless, we will assume that the dominant contribution to the perturbations in the matter sector is mainly due to the inhomogeneities of the inflaton field. In other words, \textbf{$\delta T_{ab}^{\text{rad}}$ should  be negligible compared to  $\delta T_{ab}^{\text{inf}}$}. We remind the reader that, at this point, we are not indicating that there are inhomogeneities of any definite size in the Universe, but merely one is considering what would be the dynamics of any such small inhomogeneity if it existed. The issue of their presence and magnitude is dealt with at the quantum level; as a matter of fact, if there has been no collapse of the wave function at this point, $\delta T_{ab} = \bra 0| \delta \hat{T}_{ab}^{\text{inf}}|0 \ket + \bra 0|\delta \hat{T}_{ab}^{\text{rad}}|0 \ket = 0$, consequently $\delta G_{ab} = 0$, and the space-time is perfectly homogeneous and isotropic, it is only after the collapse that generically $\bra \Theta| \delta \hat{T}_{ab}^{\text{inf}}|\Theta \ket \neq 0$ and $ \bra \Theta |\delta \hat{T}_{ab}^{\text{rad}}|\Theta \ket \neq 0$, thus, $\delta T_{ab} \neq 0$. This will be made more clear in the next section. For now, we will just continue with the classical analysis and deal with the quantum treatment in the next section.

Einstein's equations for the background  $G_{00}^{(0)}=8\pi G T_{00}^{(0)}=8\pi G a^2 \rho$   yield  Friedmann's equations. Since we are assuming that the Universe is dominated by radiation, the energy contribution of the inflaton to the total energy density $\rho$ will be negligible; therefore, the equation of state is to a good approximation $P=\rho/3$. Given the previous equation of state, one can find the explicit expression for the scale factor, this is

\beq\label{aeta}
a(\eta) = C(\eta-\eta_r)+ a_r,
\eeq
where $\eta$ is the conformal time, $C$ is a constant, $\eta_r$ is the conformal time at the beginning of the radiation era and $a_r=a(\eta_r)$. Normalizing the scale factor today as $a_0 = 1$ and by assuming that inflation ends at an energy scale of $10^{15}$ GeV, one can find the numerical values $\eta_r \simeq -1.2 \times 10^{-22}$ Mpc, $a_r \simeq 2.4 \times 10^{-28}$ and $C \simeq 1.6 \times 10^{-6}$ Mpc$^{-1}$.

Furthermore, we will ignore for the most part of the treatment the reheating era. In other words, we will assume that the inflationary regime ends at a conformal time $\eta_{ei} \simeq -10^{-22}$ Mpc and for all practical purposes $\eta_{ei} \simeq \eta_r$. 

Now we will focus on the perturbations. The perturbed space-time will be represented by the line element

\beq\label{metric}
ds^2 = a(\eta)^2 [-(1+2\Phi) d\eta^2 + (1-2\Psi) \delta_{ij} dx^idx^j],
\eeq
where  we have focused only on the scalar perturbations and have chosen to work in the longitudinal gauge. 

As we have said, the contribution from $\delta T_{ab}^{\text{rad}}$ to the perturbations of the matter sector  is negligible compared to $\delta T_{ab}^{\text{inf}}$. Thus,

\beq
\delta G_{ab} = 8 \pi G \delta T_{ab}^{\text{inf}}.
\eeq
Furthermore we can write the scalar field as as follows: $\phi (\x,\eta) = \phi_0 (\eta) + \dphi (\x,\eta)$, where $\dphi \ll \phi_0$. 

Einstein's equations at first order in the perturbations,  $\delta G_0^0 = 8 \pi G \delta T_0^0$, $\delta G_i^0 = 8 \pi G \delta T_i^0$ and $\delta G^i_j = 8 \pi G \delta T^i_j$, are given respectively by

\beq\label{00inf1}
\nabla^2 \Psi -3\mH(\mH\Phi + \Psi') = 4 \pi G [-\phi_0'^2 \Phi + \phi_0' \dphi' + \partial_\phi V a^2 \dphi],
\eeq

\beq\label{0iinf1}
\partial_i (\mH \Phi + \Psi') = 4 \pi G \partial_i ( \phi_0' \dphi),
\eeq

\barr\label{ijinf1}
 [\Psi'' + \mH(2\Psi+\Phi)' + (2\mH' + \mH^2)\Phi &+& \textstyle{\half} \nabla^2 (\Phi - \Psi)] \delta^i_j - \textstyle{\half} \partial^i \partial_j (\Phi - \Psi) = \nonumber \\
& &  4 \pi G [\phi_0' \dphi' -\phi_0'^2 \Phi  - \partial_\phi V a^2 \dphi]\delta^i_j.
\earr
It  is  easy to see that  for the case $i\not=j$ in Eq. \eqref{ijinf1}, together with appropriate  boundary conditions (more easily seen in the Fourier transformed  version), leads to  $\Psi = \Phi$; from now on we will use this result. 

By combining Eqs. \eqref{00inf1} and \eqref{0iinf1}, one obtains

\beq\label{master0}
\nabla^2 \Psi + 4\pi G \phi_0'^2 \Psi = 4 \pi G [ \phi_0' \delta \phi' + (a^2 \partial_\phi V + 3 \mH \phi_0') \delta \phi].
\eeq

After decomposing $\Psi$ and $\phi$ in Fourier modes, the above equation yields

\beq\label{master1}
\Psi_{\nk} (\eta) = \frac{ 4 \pi G \phi_0' (\eta) }{-k^2 + 4\pi G \phi_0' (\eta)^2} \left[ \dphi'_{\nk} (\eta) + \left( 3\mH + \frac{a^2 \partial_\phi V }{\phi_0'(\eta)}  \right)  \dphi_{\nk} (\eta)\right].  
\eeq

The energy density of the scalar field is $\rho_\phi = T_{00}^{\text{inf}}$. Since the Universe is radiation dominated and the inflationary era has ended,  the scalar field is now rapidly oscillating around the minimum of its potential, this is  $\partial_\phi V \simeq 0$; therefore, we can approximate the energy density of the inflaton as $\rho_\phi \simeq \phi_0'^2/2a^2 \ll \rho_{\text{rad}}$. Thus, Eq. \eqref{master1} is rewritten as

\beq\label{master2}
\Psi_{\nk} (\eta) = \frac{\sqrt{\rho_\phi}}{\sqrt{2} M_P^2 \left( -k^2 + \rho_\phi a^2 / M_P^2 \right)} \left[ a \dphi'_{\nk} (\eta) + 3 \mH a \dphi_{\nk} (\eta) \right],
\eeq
where we used the definition of the reduced Planck's mass $M_P^2 \equiv (8\pi G)^{-1}$. Equation \eqref{master2} relates the perturbations in the inflaton field with the perturbations of the metric. 

Moreover, Eq. \eqref{master2} was obtained by combining Eqs. \eqref{00inf1} and \eqref{0iinf1} which correspond to Einstein's equations with components $\delta G^{0}_0 = 8\pi G \delta T^0_0$ and $\delta G^{0}_i = 8\pi G \delta T^0_i$; it is a well known result \cite{Wald84} that these particular equations are not actual motion equations but rather constraint equations. The motion equation is the one given by $\delta G^i_j = 8\pi G \delta T^i_j$ [Eq. \eqref{ijinf1}],  from this equation (with $i = j$)  one can derive the metric perturbation motion equation; for the epoch corresponding to a radiation-dominated  Universe, the motion equation for the modes $\Psi_k$ takes the form

\beq\label{movpsi}
\Psi_{\nk}'' (\eta) + \frac{4}{\eta-\eta_r + a_r/C} \Psi_{\nk}' (\eta)  + \frac{k^2}{3} \Psi_{\nk} (\eta) = 0
\eeq
The analytical solution to Eq. \eqref{movpsi} is:

\barr\label{psisolucion}
\Psi_{\nk} (\eta) &=& \frac{3}{(k\eta-\delta_k)^2} \bigg\{ C_1 (\nk) \left[ \frac{\sqrt{3}}{k\eta-\delta_k} \sin \left( \frac{k\eta-\delta_k}{\sqrt{3}}  \right) - \cos \left( \frac{k\eta-\delta_k}{\sqrt{3}}  \right)   \right]  \nonumber \\
&+&  C_2 (\nk) \left[ \frac{\sqrt{3}}{k\eta-\delta_k} \cos \left( \frac{k\eta-\delta_k}{\sqrt{3}}  \right) + \sin \left( \frac{k\eta-\delta_k}{\sqrt{3}}  \right)   \right] \bigg\},
\earr
with $\delta_k \equiv k\eta_r - k a_r/C$.  Once the collapse has created all modes $\Psi_k$ (as will be argued in more detail in Sec. \ref{semiclassical}), we can divide them in two types: 

\begin{itemize}
\item Modes with an associated proper wavelength bigger than the Hubble radius, we will call these the super-horizon modes.

\item Modes with an associated proper wavelength smaller than the Hubble radius, we will call these the sub-horizon modes.\footnote{The condition that modes are smaller than the horizon is given by $k \gg aH = \mH$, by using the exact expression for $\mH$ during the radiation dominated epoch $\mH \equiv a'(\eta)/a(\eta) =  1/(\eta-\eta_r + a_r/C)$, one checks that the latter condition is equivalent to $(k\eta-\delta_k) \gg 1$. Alternatively, modes that are super-horizon during radiation satisfy $(k\eta-\delta_k)  \ll 1 $.}  
\end{itemize}

If $(k\eta-\delta_k) \gg 1$ the general solution, Eq. \eqref{psisolucion}, approaches zero; in other words, for sub-horizon modes $\Psi_{\nk} \to 0$. On the other hand, the dynamics of the super-horizon modes, i.e. those that satisfy $(k\eta-\delta_k) \ll 1$, is given by

\beq
\Psi_{\nk} (\eta) = \frac{C_1 (\nk)}{3} + \frac{3^{3/2} C_2 (\nk) }{(k\eta-\delta_k)^3}.
\eeq
The second mode is known as the decaying mode which we shall neglect hereafter. Since sub-horizon modes decay as $1/(k\eta-\delta_k)^2 \propto 1/a(\eta)^2$, they cannot account for the modes of interest in the angular power spectrum; conversely, super-horizon modes are constant until they enter the horizon. Therefore, we will only focus on super-horizon modes

\beq\label{psisuperhor}
\Psi_{\nk} (\eta) \simeq \frac{C_1 (\nk)}{3}.
\eeq
The constant $C_1(k)$ can be obtained from Eq. \eqref{master2}, which, as we said, corresponds to a constraint equation,  evaluated at some particular time, say $\tc$ (later in the paper we will argue in more detail that this corresponds to the time of collapse), before the modes enters the horizon; thus,

\beq\label{masterx}
\Psi_{\nk} \simeq \frac{\sqrt{\rho_\phi}}{\sqrt{2} M_P^2 \left( -k^2 + \rho_\phi a^2 / M_P^2 \right)} \left[ a \dphi'_{\nk} (\eta) + 3 \mH a \dphi_{\nk} (\eta) \right] \bigg|_{\eta=\tc} \qquad \textrm{with} \qquad k\tc-\delta_k \ll 1.
\eeq

We want to emphasize that at this point the analysis has been done in a classical manner, the quantum aspects will be analyzed in the next section. Nevertheless, we have shown that  the super-horizon modes for the curvature perturbation are constant during the radiation era, if $\Psi_{\nk}$ is classical, and, thus, follows a dynamical evolution given by Einstein's (classical) equations.


\section{Quantum analysis of the perturbations}\label{quantum}

In this section we proceed to establish the quantum theory of the inflaton perturbations. The difference with previous works \cite{PSS06,US08,DT11,Leon11} is that, in the case of the present work, the scale factor of the background metric is given by Eq. \eqref{aeta}, which corresponds to a radiation dominated Universe; while in the cited works, the scale factor corresponds to a (quasi) de-Sitter type of Universe. Consequently, we will construct the quantum theory of a scalar field in a radiation FRW background Universe. 

We start  by writing the action:

\beq\label{actioncol2}
S_{\text{inf}} =  \int d^4x \sqrt{-g} \bigg[  -\half \nabla_a \phi \nabla_b \phi g^{ab} - V[\phi] \bigg].
\eeq

 Our fundamental quantum variable will be the fluctuation of the inflaton field, $\dphi (\x,\eta)$; however, it will be easier to work  with the rescaled field variable $y=a\dphi$. Next we expand the action \eqref{actioncol2} up to second order in the rescaled variable (i.e. up to second order in the scalar field fluctuations)

\beq\label{acciony}
\delta S^{(2)}= \int d^4x \delta \mathcal{L}^{(2)} = \int d^4x \frac{1}{2} \left[ y'^2 - (\nabla y)^2 + \left(\frac{a'}{a} \right)^2 y^2 - 2 \left(\frac{a'}{a} \right) y y' \right].
\eeq
The canonical momentum conjugated to $y$ is $\pi \equiv \partial \delta \mathcal{L}^{(2)}/\partial y' = y'-(a'/a)y=a\dphi'$. The field and momentum variables are promoted to operators satisfying the equal time commutator relations $[\hat{y}(\x,\eta), \hat{\pi}(\x',\eta)] = i\delta (\x-\x')$ and $[\hat{y}(\x,\eta), \hat{y}(\x',\eta)] = [\hat{\pi}(\x,\eta), \hat{\pi}(\x',\eta)] = 0$. We expand the momentum and field operators in Fourier modes

\beq
\hat{y}(\eta,\x) = \frac{1}{L^3} \sum_{\nk} \hat{y}_{\nk} (\eta) e^{i \nk \cdot \x}  \qquad \hat{\pi}(\eta,\x) = \frac{1}{L^3} \sum_{\nk} \hat{\pi}_{\nk} (\eta) e^{i \nk \cdot \x},
\eeq
where the sum is over the wave vectors $\vec k$ satisfying $k_i L=2\pi n_i$ for $i=1,2,3$ with $n_i$ integer and $\hat y_{\nk} (\eta) \equiv y_k(\eta) \ann_{\nk} + y_k^*(\eta) \cre_{-\nk}$ and  $\hat \pi_{\nk} (\eta) \equiv g_k(\eta) \ann_{\nk} + g_{k}^*(\eta) \cre_{-\nk}$. From the previous expression it is clear that we are taking the quantization on a finite cubic box of length $L$, at the end of the calculations we will go to the continuum limit ($L \to \infty$, $k \to $ cont.). The equation of motion for  $y_k(\eta)$ derived from action \eqref{acciony} is

\beq\label{ykmov}
y_k''+\left( k^2 - \frac{a''}{a} \right) y_k = 0.
\eeq

It is worthwhile to mention that the scale factor $a$ corresponds to the radiation dominated era. In such case, the scale factor is given as in Eq. \eqref{aeta}, consequently the motion equation \eqref{ykmov} is written as

\beq
y_k'' + k^2 y_k = 0,
\eeq
which is the motion equation of a harmonic oscillator. The solutions are, thus,

\begin{subequations}\label{modosrad}
\beq
y_k(\eta) = A_k e^{ik\eta} + B_k e^{-ik\eta},
\eeq
\beq
g_k (\eta) = -A_k k \left( \frac{\mH}{k} - i \right) e^{ik\eta} - B_k k \left( \frac{\mH}{k} + i \right) e^{ik\eta},
\eeq
\end{subequations}
where $A_k$ and $B_k$ are constants that are fixed by the canonical commutation relations between $\hat y$ and $\hat \pi$, which give $[\hat{a}_{\nk},\hat{a}^\dag_{\nk'}] = L^3 \delta_{\nk,\nk'}$, thus $y_k(\eta)$ must satisfy $y_k g_k^* - y_k^* g_k = i$ for all $k$ at some time $\eta$; however, this condition alone does not completely fix the constants $A_k$ and $B_k$. One still needs to select a choice for the vacuum state for the field. In order to proceed, we will select a vacuum state in the inflation era (where $a''/a \simeq 2 \eta^{-2}$), where the quantum fluctuations of the inflaton field are originated.  There are a variety of choices regarding the vacuum state during inflation, one of the most common choices is the so-called Bunch-Davies (BD) vacuum characterized by

\beq\label{ykbd}
y_k (\eta) = \frac{1}{\sqrt{2k}} \left(1 - \frac{i}{k\eta} \right) e^{-ik\eta}, \qquad g_k (\eta) = -i \sqrt{\frac{k}{2}} e^{-ik\eta}.
\eeq

Consequently, the constants $A_k$ and $B_k$ will be fixed by matching the modes during the inflation [Eqs. \eqref{ykbd}] era  and the modes during the radiation era [Eqs. \eqref{modosrad}]  at the time $\eta_r$,  which corresponds to the conformal time of the beginning of the radiation era and is essentially the same order of magnitude as the conformal time that marks the end of inflation. Note that we are neglecting the reheating era that describes the decay of the inflaton in all the fields characterizing the radiation type of matter. If one takes into account the interaction of the inflaton and the quantum fields representing the radiation matter, the vacuum state could possibly change; however,  such new vacuum state would still be perfectly homogeneous and isotropic. In other words, the reheating period cannot break the symmetry of an original quantum state because its dynamics is given by the Schroedinger's equation which preserves the symmetry. For simplicity we will not consider the reheating period and assume that all the fields, before and after inflation, are characterized by the BD vacuum state.

Therefore, with the previous assumptions,  the constants $A_k$ and $B_k$ are

\beq\label{AkBk}
A_k = \frac{e^{-2ik\eta_r}}{2^{3/2}k^{5/2}\eta_r^2}, \qquad B_k = \frac{1}{\sqrt{2k}} \left( 1 - \frac{i}{k\eta_r} \right) - \frac{1}{2^{3/2}k^{5/2}\eta_r^2}.
\eeq

To recapitulate, the modes $y_k(\eta)$ are originated during the inflationary epoch in the BD vacuum state, after inflation reaches its end at $\eta_r$ (and ignoring the reheating era), the radiation dominated epoch begins and the inflaton is now oscillating around the minimum of its potential. Additionally, its modes continue to evolve according to Eqs. \eqref{modosrad};  nevertheless, the quantum state of the modes is still the BD vacuum state, which is 100\% homogeneous and isotropic; consequently there are no inhomogeneities and anisotropies present at this stage of the evolution. Thus, as discussed in Sec. \ref{intro},  in order to account for the issue regarding the emergence of an anisotropic and inhomogeneous Universe from an exactly isotropic and homogeneous initial state of the primordial perturbations, we must consider a self-induced collapse of  the wave function. In the following section, we will describe how to parameterize such collapse and show how the primordial curvature perturbations are produced by the self-induced collapse in a radiation dominated era. 

\section{The collapse model and the curvature perturbation }
\label{collapseschemes}

In this section, we will show how one can generate the primordial curvature perturbation during the radiation dominated era by introducing the collapse hypothesis. 

The self-induced collapse hypothesis is based on considering that the collapse acts similar to a ``measurement'' (clearly, there is no external observer or detector involved). This lead us to consider Hermitian operators, which in ordinary quantum mechanics are the ones susceptible of direct measurement. Therefore, we separate $\hat y_{\nk} (\eta)$ and $\hat \pi_{\nk} (\eta)$ into their real and imaginary parts $\hat y_{\nk} (\eta)=\hat y_{\nk}{}^R (\eta) +i \hat y_{\nk}{}^I (\eta)$ and $\hat \pi_{\nk} (\eta) =\hat \pi_{\nk}{}^R (\eta) +i \hat \pi_{\nk}{}^I (\eta)$ in this way the operators $\hat y_{\nk}^{R, I} (\eta)$ and $\hat \pi_{\nk}^{R, I} (\eta)$ are  hermitian operators. Thus,

\beq\label{operadoresRI}
\hat{y}_{\nk}^{R,I} (\eta) = \sqrt{2} \mathcal{R}[y_k(\eta) \hat{a}_{\nk}^{R,I}], \qquad \hat{\pi}_{\nk}^{R,I} (\eta) = \sqrt{2} \mathcal{R}[g_k(\eta) \hat{a}_{\nk}^{R,I}],
\eeq
where $\hat{a}_{\nk}^R \equiv (\hat{a}_{\nk} + \hat{a}_{-\nk})/\sqrt{2}$, $\hat{a}_{\nk}^I \equiv -i (\hat{a}_{\nk} - \hat{a}_{-\nk})/\sqrt{2}$. The commutation relations for the $\hat{a}_{\nk}^{R,I}$ are non-standard

\beq\label{creanRI}
[\hat{a}_{\nk}^R,\hat{a}_{\nk'}^{R \dag}] = L^3 (\delta_{\nk,\nk'} + \delta_{\nk,-\nk'}), \quad [\hat{a}_{\nk}^I,\hat{a}_{\nk'}^{I \dag}] = L^3 (\delta_{\nk,\nk'} - \delta_{\nk,-\nk'}),
\eeq
with all other commutators vanishing.

One natural way to proceed is to assume that the effect of the collapse on a state is  analogous  to some  sort of  approximate measurement; in other words,   after the  collapse, the expectation values of the field and momentum operators  in each mode  will  be related to the uncertainties  of the  initial  state. In the vacuum state, $\hat{y}_{\nk}$ and $\hat{\pi}_{\nk}$ are individually distributed according to Gaussian wave functions centered at 0 with spread $(\Delta \hat{y}_{\nk})^2_0$ and $(\Delta\hat{\pi}_{\nk})^2_0$, respectively. We   consider  various possibilities  for such relations; we will refer  to them as ``collapse schemes'' to the different ways of characterizing the expectation values.  So, even though we did not assume a specific collapse mechanism, the different schemes refer to different ways of the collapse to happen, affecting either the field or momentum variable or both. The most generic form to characterize such ``collapse schemes'' is

\begin{subequations}\label{esquemas}
\beq
 \bra \hat{y}^{R,I}_{\nk}(\eta^c_{\nk})\ket_{\Theta}  = \lambda_1 x_{\nk,1}^{R,I}
  \sqrt{\left(\Delta \hat{y}^{R,I}_{\nk} (\eta_k^c) \right)^2_0} = \lambda_1 x_{\nk,1}^{R,I} \frac{L^{3/2}}{\sqrt{2}} |y_k (\tc)|, 
  \eeq
  \beq
  \bra \hat{\pi}^{R,I}_{\nk}(\eta^c_k) \ket_{\Theta} = \lambda_2 x_{\nk,2}^{R,I}
  \sqrt{\left(\Delta \hat{\pi}^{R,I}_{\nk} (\eta_k^c) \right)^2_0} = \lambda_2 x_{\nk,2}^{R,I} \frac{L^{3/2}}{\sqrt{2}} |g_k (\tc)|.
\eeq
\end{subequations}
The subindex $\bra \cdot \ket_{\Theta}$ represents that we are taking the expectation value on the post-collapse state $|\Theta \ket$. The random variables $x_{\nk,1}^{R,I}, x_{\nk,2}^{R,I}$ are distributed according to a Gaussian centered at zero, of spread one (normalized), and are statistically uncorrelated; the quantity $\tc$ denotes the conformal time of collapse, which in principle might depend on $k$. The parameters $\lambda_1, \lambda_2$ can only take two values: 0 or 1, the only purpose of these  parameters is to  ``switch on'' or ``switch off'' the operators in which the collapse take place. For example, we can choose a scheme in which the momentum operator is affected by the collapse but not the field, i.e. $\bra \hat{\pi}_{\nk} (\tc) \ket_{\Theta} \neq 0$, $\bra \hat{y}_{\nk} (\tc) \ket_{\Theta} = 0$, this situation corresponds to set $\lambda_2=1$, $\lambda_1=0$. In section \ref{analisis} we will study with detail the primordial spectrum in three different cases: i) only the field variable is affected by the collapse, $\lambda_1=1$, $\lambda_2=0$; ii) only the momentum  variable is affected by the collapse, $\lambda_1=0$, $\lambda_2=1$; iii) both variables are affected by the collapse, $\lambda_1=1$, $\lambda_2=1$.

The next step would be to relate the quantum objects with the observational quantities, but before we proceed in that direction, we will like to introduce the way in which we believe the quantum degrees of freedom (DOF) relate to the classical description of the space-time in terms of the metric.

\subsection{The semiclassical gravity approach and the collapse of the wave function}\label{semiclassical}

We will rely on the so-called ``semiclassical gravity'' approach. This approach is characterized by Einstein's semiclassical equations $G_{ab} = 8 \pi G \bra \hat{T}_{ab} \ket$, which relate the matter quantum DOF with the classical description of gravity in terms of the metric. The semiclassical approach is a valid approximation in the energy scales for our case of interest, also, this approach lead us to   consider that the Universe can be  described,  by what was called  \textit{Semiclassical Self-consistent Configuration} (SSC), first introduced in Ref. \cite{DT11}; in the following, we present a brief description of such idea.


The SSC considers  a space-time  geometry characterized by a  classical  space-time  metric  and  a standard quantum field theory  constructed  on that  fixed space-time background, together with a particular quantum  state in that construction such that the  semiclassical   Einstein's  equations hold. Specifically, one will establish that the set 

\beq
\left\lbrace g_{\mu\nu}(x),\hat{\varphi}(x),\hat{\pi}(x),\mathscr{H},\vert\xi\rangle\in\mathscr{H}\right\rbrace
\eeq
characterizes a SSC if and only if $\hat{\varphi}(x)$, $\hat{\pi}(x)$ and $\mathscr{H}$  correspond to a quantum field theory constructed over a space-time  with  metric  $g_{\mu\nu} (x)$ (as  described in, say \cite{Wald94}), and  the  state $\vert\xi\rangle$ in $\mathscr{H}$  is  such that

\begin{equation}\label{Mset-up}
G_{\mu\nu}[g(x)]=8\pi G\langle\xi\vert \hat{T}_{\mu\nu}[g(x),\hat{\varphi}(x),\hat{\pi}(x)]\vert\xi\rangle,
\end{equation}
for all the points in the space-time manifold.

Such  description  is  thought to  be  appropriate  in the   regime of interests  except  in those  times  when  a  collapse  occurs. In particular, if one considers a specific collapse mechanism, then Eq. \eqref{Mset-up} will not hold; this is due to the fact that the quantum collapse would induce sudden changes or ``state jumps'' to the initial quantum state, thus the divergence $\nabla_a \bra \hat{T}^{ab} \ket \neq 0$ which implies that $\nabla_a G^{ab} \neq 0$; evidently that is a problem since  a well-known result from General Relativity  is that the divergence of  Einstein's tensor vanishes.  Nevertheless,  since we will be only interested in states \emph{before} and \emph{after} the collapse, this breakdown of the semiclassical approximation would not be important for our present work. During the collapse, the dynamics of the space-time would be affected, but in the absence of a full workable theory of quantum gravity, we cannot characterize the metric dynamical response to the modification of the standard unitary quantum evolution.

The relation between the SSC and the collapse process can be described in a more formal way: first, within the Hilbert space associated to the  given SSC-i, one can consider that a transition $\vert\xi^{\textrm{(i)}}\rangle\to\vert\zeta^{\textrm{(i)}}\rangle_{\textrm{target}}$ ``is about to happen", with both $\vert\xi^{\textrm{(i)}}\rangle$ and $\vert\zeta^{\textrm{(i)}}\rangle_{\textrm{target}}$ in $\mathscr{H}^{\textrm{(i)}}$. In general, the set $\{g^{\textrm{(i)}},\hat{\varphi}^{\textrm{(i)}},\hat{\pi}^{\textrm{(i)}}, \mathscr{H}^{\textrm{(i)}},\vert\zeta^{\textrm{(i)}}\rangle_{\textrm{target}}\}$ will not characterize a new SSC. In order to describe a reasonable picture,  as presented in Ref.  \cite{DT11},  one needs to relate  the  state $\vert\zeta^{\textrm{(i)}}\rangle_{\textrm{target}}$ with another one $\vert\zeta^{\textrm{(ii)}}\rangle$ ``existing'' in a new Hilbert space $\mathscr{H}^{\textrm{(ii)}}$ for which $\{g^{\textrm{(ii)}},\hat{\varphi}^{\textrm{(ii)}},\hat{\pi}^{\textrm{(ii)}}, \mathscr{H}^{\textrm{(ii)}},\vert\zeta^{\textrm{(ii)}}\rangle\}$ is a valid SSC; this  new SSC is denoted by SSC-ii. Consequently, one needs to determine first the ``target'' (non-physical) state in $\mathscr{H}^{\textrm{(i)}}$ to which the initial state is ``tempted'' to jump, sort of speak, and after that, one can relate such target state with a corresponding  state  in the Hilbert space of a new SSC, the SSC-ii. One then considers that the target state is chosen stochastically, guided by the quantum uncertainties of designated  field  operators, evaluated on the initial state $\vert\xi^{\textrm{(i)}}\rangle$,  at the collapsing time; this was the motivation behind the characterization of the collapse schemes presented in  Eqs. \eqref{esquemas}.

Regarding the identification between the two different SSC's involved in the collapse, the prescription introduced in Ref. \cite{DT11} is the following: Assume that the collapse takes place along a Cauchy hyper-surface $\Sigma$. A transition from the physical state $\vert\xi^{\textrm{(i)}}\rangle$ in $\mathscr{H}^{\textrm{(i)}}$ to the physical state $\vert\zeta^{\textrm{(ii)}}\rangle$ in $\mathscr{H}^{\textrm{(ii)}}$ (associated to a certain  target \textit{non-physical} state $\vert\zeta^{\textrm{(i)}}\rangle_{\textrm{target}}$ in $\mathscr{H}^{\textrm{(i)}}$) will occur  in a way that

 \begin{equation}\label{recipe.collapses}
_\textrm{target}\langle\zeta^{\textrm{(i)}}\vert \hat{T}^{\textrm{(i)}}_{\mu\nu}[g^{\textrm{(i)}}, \hat{\varphi}^{\textrm{(i)}},\hat{\pi}^{\textrm{(i)}}]\vert\zeta^{\textrm{(i)}}\rangle_{\textrm{target}} \big|_{\Sigma}= \langle\zeta^{\textrm{(ii)}}\vert \hat{T}^{\textrm{(ii)}}_{\mu\nu}[g^{\textrm{(ii)}}, \hat{\varphi}^{\textrm{(ii)}},\hat{\pi}^{\textrm{(ii)}}]\vert\zeta^{\textrm{(ii)}}\rangle \big|_{\Sigma}\, 
\end{equation}
i.e.  in such a  way  that  the expectation value of the energy momentum tensor, associated to the states $\vert\zeta^{\textrm{(i)}}\rangle_{\textrm{target}}$ and $\vert\zeta^{\textrm{(ii)}}\rangle$ evaluated on the Cauchy hyper-surface $\Sigma$, coincides. Note that the left hand side  in the expression  above is meant  to be constructed from the elements of the SSC-i (although $\vert\zeta^{\textrm{(i)}}\rangle_{\textrm{target}}$ is not really {\it the state} of the SSC-i), while the right hand side correspond to quantities  evaluated  using the SSC-ii.

In the situation of interest for this work, the SSC-i corresponds to a homogeneous and isotropic space-time characterized by $\Psi=0$ with the state of the quantum field corresponding to the Bunch-Davies vacuum. Meanwhile, the SSC-ii corresponds to an excitation of all the modes $k$, characterized by the Newtonian potential $\Psi_{\nk}$. In particular, the post-collapse state $|\zeta^{\textrm{(ii)}} \ket$  is explicitly

\beq
|\zeta^{\textrm{(ii)}} \ket = \ldots  |\zeta^{\textrm{(ii)}}_{-\nk_2}  \ket \otimes |\zeta^{\textrm{(ii)}}_{-\nk_1} \ket \otimes |\zeta^{\textrm{(ii)}}_0 \ket \otimes |\zeta^{\textrm{(ii)}}_{\nk_1} \ket \otimes |\zeta^{\textrm{(ii)}}_{\nk_2} \ket \ldots,
\eeq
which means that the collapse process affects all modes of the quantum field. Given the previous prescription for the post-collapse state, and considering the SSC-ii, we can now associate each mode of the post-collapse state to each mode characterized by $\Psi_{\nk}$. In this way the metric perturbations $\Psi(x)$ are born, and, thus  the SSC-ii, corresponds to an inhomogeneous and anisotropic space-time at all scales $k$; in particular,  $\Psi_{\nk}$ corresponds to modes that are super-horizon and sub-horizon.

One advantage of relying on the semiclassical approach is that it allows to present a clear picture of the physical process (although not exactly known) responsible for the birth of the primordial perturbations from the quantum collapse: the initial state of the Universe is described by both an homogeneous-isotropic vacuum state and an equally homogeneous-isotropic Friedmann-Robertson-Walker space-time. Then, at some point during the radiation epoch, some unknown physical mechanism,  causes  a quantum collapse of the matter field wave function. However, the state resulting from the collapse needs not to share the same symmetries as the initial state. After the collapse, the gravitational DOF are assumed to be, once more, accurately described by Einstein's semiclassical equation. Nevertheless,  $\bra \hat{T}_{ab} \ket$ evaluated in the new state does not generically posses the symmetries of the pre-collapse state; hence, we are led to a new geometry that is no longer homogeneous and isotropic.

We should note  here that  we  will   not be using  at this point the  full  fledged  formal  treatment developed. This is because, as  can be  see in Ref. \cite{DT11},  the  problem  becomes extremely cumbersome even in  the treatment of a single  mode. Thus, even though it  is in principle  possible to use  such  detailed  formalism   to treat   the  complete  set of relevant modes, when studying the  CMB spectrum  the  task  quickly becomes a practical impossibility.  We  will instead rely on the  less formal  treatments  we had  employed in previous works. This is, we can assume that after the collapse has ended, and having  constructed a SSC-ii, we can generalize  Eq. \eqref{masterx} in the following manner:

\beq\label{master4}
\Psi_{\nk} (\tc) = \frac{\sqrt{\rho_\phi}}{\sqrt{2} M_P^2 \left( -k^2 + \rho_\phi a_c^2 / M_P^2 \right)} \bigg( \bra\hat{\pi}_{\nk} (\tc)  \ket + 3 \mH_c \bra\hat{y}_{\nk} (\tc)  \ket  \bigg)  ,
\eeq
with $a_c \equiv a(\tc)$ and $ \mH_c \equiv \mH (\tc) $.  The condition that the associated proper wavelength of the modes is bigger than the Hubble radius at the time of collapse is given by $k \tc - \delta_k \ll 1$; but upon using the numerical values for $a_r,\eta_r,C$ one obtains that $\delta_k \simeq 10^{-22}$, thus, the time of collapse must satisfy $k\tc \ll 1$. 

Equation \eqref{master4} is the main result of this section as it relates the primordial curvature perturbation with the quantum expectation values after the collapse; i.e. is an expression that relates the metric perturbation with the parameters characterizing the collapse. In this manner, the quantum collapse of the wave function can generate the primordial cosmic seeds at the radiation era. Note that, as discussed above, the collapse affects all modes, therefore we could use Eq. \eqref{masterx}, which corresponds to the super-horizon modes. The sub-horizon modes are present too, but as shown in Sec. \ref{classical}, they decay as $1/a(\eta)^2$. Furthermore, within the semiclassical approach, the metric is always a classical object, therefore its dynamics during the radiation era, is exactly given by the motion equation \eqref{movpsi}, and as we have argued, it will not be modified once the collapse mechanism has ended.

It is worth noting that, by relying on the semiclassical approach, we have no issue regarding the ``quantum-to-classical'' transition that is always present in the traditional approach, namely, to find a justification from going from an strictly quantum object $\hat{\Psi}_{\nk}$ to a classical stochastic field $\Psi_{\nk}$. The next task is to obtain an equivalent power spectrum for the primordial perturbations that can be consistent with the observational data.

Regarding the tensor modes and the semiclassical gravity approach, we should mention that recent observational data \cite{BICEP2} suggest that the amplitude corresponding to the tensor modes may be non-trivial. Additionally, in our approach, the source of the curvature perturbations lies in the quantum inhomogeneities of the inflaton field (after the collapse). Once the collapse has taken place, the inhomogeneities of the inflaton feed into the gravitational DOF leading to perturbations in the metric components. However, the metric itself is not a source of the self-induced collapse. Therefore, as the scalar field does not act as a source for the metric tensor modes, at least not at first-order considered here, the analysis concerning the amplitude of the primordial gravitational waves should be done at second-order in the perturbations; such analysis is beyond the scope of this paper and would be the subject of future research. On the other hand, if one takes the view that both, metric and matter perturbations should be quantized, say at the level of the Mukhanov-Sasaki variable, then one could still implement a specific collapse mechanism for this variable. Furthermore, quantizing matter and metric perturbations would yield a non-trivial amplitude for first-order tensor modes (in the same vein as in the standard approach), after putting into effect a mechanism responsible for collapsing the wave function, one can look for possible modifications to tensor power spectra and their implications. In the particular case of the CSL mechanism, this type of analysis has been done in Ref. \cite{TPsinghBICEP2}.

\section{Observational quantities}
\label{oquantities}

In this section, we will relate the parameters characterizing the collapse with the observational quantities.

The temperature anisotropies $ \frac{\delta T}{T_0}$ of the CMB are clearly the most direct observational quantity available ($T_0$ is the mean temperature). One can expand such anisotropies with the help of the spherical harmonics  $ \frac{\delta T}{T_0} (\theta,\varphi) = \sum_{l,m} a_{lm} Y_{lm} (\theta,\varphi)$; therefore, the coefficients $a_{lm}$ are given by

\beq\label{alm0}
a_{lm} = \int \Theta (\hat n) Y_{lm}^\star (\theta,\varphi) d\Omega,
 \eeq
with $\hat n = (\sin \theta \sin \varphi, \sin \theta \cos \varphi, \cos \theta)$ and $\theta,\varphi$ the coordinates on the celestial two-sphere; we have also defined $\Theta (\hat {n}) \equiv \delta T (\hat n)/ T_0$. Assuming instantaneous recombination, the relation between the primordial perturbations and the observed CMB anisotropies is

\beq\label{mastertemp}
\Theta (\hat n) = [\Psi + \frac{1}{4} \delta_\gamma] (\eta_D) + \hat n \cdot \vec{v}_\gamma (\eta_D) + 2 \int_{\eta_D}^{\eta_0} \Psi'(\eta) d\eta, 
\eeq
where $\eta_D$ is the time of decoupling; $\delta_\gamma$ and $\vec{v}_\gamma$ are the density perturbations and velocity of the radiation fluid (which are generated after the collapse, i.e. once the curvature perturbation $\Psi$ is originated). 

It is common practice to decompose the temperature anisotropies in Fourier modes

\beq
\Theta (\hat n) = \sum_{\nk} \frac{\Theta (\nk)}{L^3} e^{i \nk \cdot R_D \hat n},
\eeq
with $R_D$ the radius of the last scattering surface. Afterwards, one solves the fluid motion equations with the initial condition $\Psi_{\nk}$, which in our model corresponds to $\Psi_{\nk} (\tc)$, i.e. the curvature perturbation at the time of collapse, Eq. \eqref{master4}.

Furthermore, using that $e^{i \nk \cdot R_D \hat n} = 4 \pi  \sum_{lm} i^l j_l (kR_D) Y_{lm} (\theta,\varphi) Y_{lm}^\star (\hat k )$,  expression \eqref{alm0} can be rewritten as

\beq\label{alm1}
a_{lm} = \frac{4 \pi i^l}{L^3} \sum_{\nk} j_l (kR_D) Y_{lm}^\star(\hat k) \Theta (\nk),
\eeq
with $j_l (kR_D)$ the spherical Bessel function of order $l$.

The linear evolution which relates the initial curvature perturbation $\Psi_{\nk}$ and the temperature anisotropies $\Theta (\nk)$ is summarized in the transfer function $T(k)$, in other words, $T(k)$ is the result of solving the fluid motion equations (for one mode)  with the initial condition provided by the curvature perturbation $\Psi_{\nk}$ and then make use of Eq. \eqref{mastertemp} to relate it  with the temperature anisotropies. Thus, $\Theta (\nk) = T(k) \Psi_{\nk}$.

Consequently, the coefficients $a_{lm}$, in terms of the modes $\Psi_{\nk} (\tc)$, are given by 
 
\beq\label{alm2}
a_{lm} = \frac{4 \pi i^l}{ L^3} \sum_{\nk} j_l (kR_D) Y_{lm}^\star(\hat k) T (k) \Psi_{\nk} (\tc),
\eeq
We emphasize that $\Psi_{\nk} (\tc)$  must correspond to the modes such that  $z_k \ll 1$, because as explained in Section \ref{classical} only the super-horizon modes are relevant in this context. Substituting Eq. \eqref{master4} and using Eqs. \eqref{esquemas} (i.e. the collapse schemes) in Eq. \eqref{alm2}, yields

\beq\label{almrandom}
a_{lm} = \frac{2 \pi i^l}{ L^{3/2}} \frac{\sqrt{\rho_\phi}}{M_P^2} \sum_{\nk} \frac{j_l (kR_D) Y_{lm}^\star (\hat k) T(k)}{(-k^2 + \rho_\phi a_c^2/M_P^2)} \bigg( \lambda_2 X_{\nk,2} |g_k (\tc)| + 3\mH_c \lambda_1 X_{\nk,1} |y_k (\tc) | \bigg),
\eeq
where $X_{\nk,i} \equiv x_{\nk,j}^R + i x_{\nk,j}^I$ ($j=1,2$).

One  key   aspect that  in  our  treatment  differs,  from those followed  in the   standard   approaches,  is  the manner in which  the  results  from the formalism  are   connected  to observations.  This  is    most clearly  exhibited  by   our result regarding the quantity  $a_{lm} $   in Eq. \eqref {almrandom}.  Despite the fact that we have in principle  a close    expression for   the quantity of interest,  we  cannot use Eq.  \eqref {almrandom}  to make  a definite prediction  because   the  expression involves the  numbers   $ X_{\nk,j}$  that correspond, as  we  indicated  before,  to   a    random    choice  ``made  by  nature''   in the context of the collapse process. The  way one  makes  predictions  is by regarding  the  sum  appearing   in Eq.   \eqref {almrandom}   as   representing a   kind of  two-dimensional  random walk,  i.e the sum   of complex numbers   depending on    random  choices  (characterized by the $ X_{\nk}$).  As  is  well known, for a random walk,  one cannot   predict the final displacement (which  would correspond to the    complex  quantity $a_{lm} $),    but  one  might estimate  the  most likely value of the  magnitude  of such  displacement.  Thus,  we focus  precisely  on the  most likely  value of $|a_{lm}|$,  which we  denote  by   $|a_{lm}|_{\text{M.L.}}$.    In order to compute  that  quantity,  we   make  use  of a   fiducial  (imaginary)  ensemble of  realizations of the random  walk  and  compute the    ensemble   average value over of the  total displacement.  Thus  we   identify:
 
 \beq\label{ML}
|a_{lm}|_{\text{M.L.}}^2 = \overline{|a_{lm}|^2}.
\eeq
 The over-line appearing denotes average over the    fiducial   ensemble  of possible  realizations, i.e. of possible outcomes of the random variables  where each outcome corresponds to a single Universe. Thus, we identify the ensemble average of possible realizations with most likely value, and this most likely value with the one characterizing our Universe.

The   estimate is  done  now  in the standard  way  in  which  one  deals   with  such  random  walks:
 
 \barr\label{ML2}
|a_{lm}|^2_{\text{M.L.}} = \overline{ |a_{lm}|^2}&=& \frac{4\pi^2 \rho_\phi}{ L^3 M_P^4}\sum_{\nk,\nk'} \frac{j_l(kR_D) j_l(k'R_D) Y_{lm}^\star (\hat k) Y_{lm} (\hat k') T(k) T(k')}{(-k^2+\rho_\phi a_c^2 /M_P^2 )(-k'^2+\rho_\phi a_c^2 /M_P^2 )} \nonumber \\
&\times& \overline{\left(  \lambda_2 X_{\nk,2} |g_k (\tc)| + 3\mH_c \lambda_1 X_{\nk,1} |y_k (\tc) | \right) \left( \lambda_2 X_{\nk',2}^\star |g_k' (\eta_{\nk'}^c)| + 3\mH_c \lambda_1 X_{\nk',1}^\star |y_k (\eta_{\nk'}^c) |   \right)},
\earr
which  upon  using the   normalized Gaussian  assumption for  fiduciary  ensemble, this is, $ \overline{X_{\nk,i} X^\star_{{\nk,j'}}} = 2 \delta_{i,j} \delta_{\nk, {\nk}'} $,  leads to

   \beq\label{ML3}
|a_{lm}|^2_{\text{M.L.}} = \frac{8\pi^2 \rho_\phi}{ L^3 M_P^4} \sum_{\nk,} \frac{  j_l (kR_D)^2 | Y_{lm} (\hat k)|^2 T(k)^2}{(-k^2 + \rho_\phi a_c^2 /M_P^2)^2} \left( \lambda_2^2 |g_k (\tc)|^2 + 9 \mH_c^2 \lambda_1^2  |y_k (\tc) | ^2 \right). 
   \eeq
   
 Finally, we can remove the   fiducial  box of  side $L$   and  pass to the continuum

   \beq\label{ML4}
|a_{lm}|^2_{\text{M.L.}} 
=  \frac{ \rho_\phi}{ \pi M_P^4} \int d^3k     \frac{  j_l (kR_D)^2 | Y_{lm} (\hat k)|^2 T(k)^2}{(-k^2 + \rho_\phi a_c^2 /M_P^2)^2} \left( \lambda_2^2 |g_k (\tc)|^2 + 9 \mH_c^2 \lambda_1^2  |y_k (\tc) | ^2 \right).
    \eeq
    
The exact expressions for $|y_k (\tc)|$ and $|g_k (\tc)|$ can be obtained from Eqs. \eqref{ykbd} [with $A_k$ and $B_k$ given in Eqs \eqref{AkBk}], these are
    
    \beq\label{yk}
    |y_k (\tc)|^2 = \frac{1}{2k} \left[ 1 + \frac{1}{2\sigma_k^4} + \frac{\cos 2D_k}{\sigma_k^2} \left(1 - \frac{1}{2\sigma_k^2} \right) - \frac{\sin 2 D_k}{\sigma_k^3}       \right]
    \eeq
    and 
    \barr\label{gk}
    |g_k (\tc)|^2 &=& \frac{k}{2} \bigg\{ \left( \frac{\mH_c^2}{k^2} +1  \right)  \left( 1 + \frac{1}{2\sigma_k^4 }    \right)     + \frac{\cos 2D_k}{\sigma_k^2} \left[ \left(    \frac{\mH_{c}^{2}}{k^2} -1  \right)  \left( 1 - \frac{1}{2\sigma_k^2 }    \right)  +\frac{2 \mH_c}{k \sigma_k}                 \right]   \nonumber \\
    &-& \frac{\sin 2 D_k}{\sigma_k^2}  \left[ - \frac{2 \mH_c}{k}\left(        1 - \frac{1}{2\sigma_k^2}     \right)     +  \left( \frac{\mH_{c}^{2}}{k^2} -1  \right) \frac{1}{\sigma_k}      \right]  \bigg\},
    \earr
    where $\sigma_k \equiv k\eta_r $, $z_k \equiv k\tc$ and $D_k \equiv z_k - \sigma_k$.

  At   this  point, one   could  focus   on   the   quantity  that   is commonly presented as a direct result from the observational data, namely
     
   \beq\label{ML6}
 C_l  \equiv \frac{1}{2l+1}  \sum_m  |a_{lm}|^2
    \eeq
 for which  we  would  have the estimate
      \barr\label{ML7}
 {C_l}^{\text{M.L.}}  &\equiv& \frac{1}{2l+1}  \sum_m  |a_{lm}|_{\text{M.L.}}^2 \nonumber \\
 &=& \frac{ \rho_\phi}{ \pi M_P^4} \int_0^\infty \frac{dk}{k}  \frac{  j_l (kR_D)^2  T(k)^2 k^3}{(-k^2 + \rho_\phi a_c^2 /M_P^2)^2} \left( \lambda_2^2 |g_k (\tc)|^2 + 9 \mH_c^2 \lambda_1^2  |y_k (\tc) | ^2 \right).
    \earr

In the standard inflationary paradigm, a well-known result is that the dimensionless power spectrum $\Delta^2 (k)$ for the curvature perturbation and the $C_l$ are related by

\beq\label{cl}
C_l = \frac{4 \pi}{9} \int_0^\infty \frac{dk}{k}  j_l^2 (kR_D) T(k)^2  \Delta^2 (k).
\eeq
 Thus, by comparing Eq. \eqref{ML7} with \eqref{cl} we can extract an ``equivalent power spectrum'' for the $\Psi_{\nk}$

\beq\label{psexacto}
\Delta^2 (k) = \frac{9 \rho_\phi}{4 \pi^2 M_P^4} \frac{k^3}{ (-k^2 + \rho_\phi a_c^2 /M_P^2)^2   } \left( \lambda_2^2 |g_k (\tc)|^2 + 9 \mH_c^2 \lambda_1^2  |y_k (\tc) | ^2 \right).
\eeq

In the next section, we will show that, under certain conditions,  the power spectrum given in Eq. \eqref{psexacto} can be approximated to yield a nearly scale invariant spectrum with the correct amplitude. 

%

\section{Analysis of the equivalent power spectrum}
\label{analisis}

In this section, we will study different cases and   show that, under specific conditions, our model reproduces a nearly flat power spectrum. In standard inflationary models, the power spectrum has a phenomenological expression: $\Delta^2 (k) = \mathcal{A} k^{n_s-1}$; with $n_s$ the scalar spectral index of the perturbations. A perfect scale-invariant spectrum corresponds to $n_s = 1$. However, the most recent results from \emph{Planck} mission rule out exact scale invariance (at over  $5 \sigma$, the spectral index is $n_s = 0.9603 \pm 0.0073$). Therefore, we will explore the conditions given in our model that lead to a nearly scale invariant spectrum. Note, however, that the departure from perfect scale-invariance will be given by having introduced the collapse hypothesis. Thus, the dependence on $k$ introduced by the collapse proposal will be different from the standard one. 

Our first approximation concerns  the scale factor at the time of collapse, namely $a_c = C(\tc - \eta_r) + a_r$; if we assume that $\tc \gg |\eta_r|$, then $a_c \simeq C\tc$; additionally  $\mH_c$ at the time of collapse is $\mH_c = (\tc-\eta_r + a_r/C  )^{-1}$, which can be approximated by  $\mH_c \simeq 1/\tc$. Thus, the power spectrum in Eq. \eqref{psexacto} is approximately 

\beq\label{ps1}
\Delta^2 (k) \simeq \frac{9 \rho_\phi}{8 \pi^2 M_P^4} \frac{k^4}{ [-k^2 + \rho_\phi (C \tc /M_P)^2]^2   } \left( \lambda_2^2 N(z_k) +  9  \lambda_1^2  M(z_k) \right),
\eeq

where

\beq
M(z_k) \equiv \frac{1}{z_k^2} \left[1 + \frac{1}{2 \sigma_k^4} + \frac{\cos(2z_k - 2\sigma_k)}{\sigma_k^2} \left( 1-\frac{1}{2\sigma_k^2}\right)  - \frac{\sin(2z_k-2\sigma_k)}{\sigma_k^3} \right]
\eeq
and

\barr
N(z_k) &\equiv&  1+ \frac{1}{z_k^2} + \frac{1}{2 \sigma_k^4} + \frac{1}{2 \sigma_k^4 z_k^2}    \nonumber \\
&+& \cos(2z_k - 2\sigma_k) \left( -\frac{1}{\sigma_k^2} + \frac{1}{z_k^2 \sigma_k^2} + \frac{1}{2 \sigma_k^4} - \frac{1}{2 z_k^2 \sigma_k^4} + \frac{2}{z_k \sigma_k^3}    \right) \nonumber \\
&-& \sin(2z_k - 2\sigma_k) \left( - \frac{2}{z_k \sigma_k^2} + \frac{1}{z_k \sigma_k^4} +\frac{1}{z_k^2 \sigma_k^3} - \frac{1}{\sigma_k^3}\right)
\earr

Moreover,  we can make another approximation by considering the fact that $\sigma_k \equiv k \eta_r \ll 1$. Hence, one can take the first two term of the series expansion for  $\sin (2\sigma_k)$ and $\cos(2\sigma_k)$  and, after performing the simplification of the terms, only retain the dominant  term, which is of order  $\mathcal{O}(\sigma_k^{-4}) $. Thus,

\beq\label{Map1}
M(z_k) \simeq \frac{1}{\sigma_k^4}  \frac{\sin^2 z_k}{z_k^2}
\eeq
and

\beq\label{Nap1}
N(z_k) \simeq \frac{1}{\sigma_k^4} \left[ \half + \frac{1}{2z_k^2} + \cos(2z_k) \left( \half - \frac{1}{2 z_k^2}\right)  -\frac{\sin(2z_k)}{z_k}  \right].
\eeq

There are two limit cases we can further analyze at this point: the limit $k^2 \ll  \rho_\phi (C \tc /M_P)^2$ or $k^2 \gg  \rho_\phi (C \tc /M_P)^2$. Let us focus on the first case. 

If $k^2 \ll  \rho_\phi (C \tc /M_P)^2$ then the power spectrum in Eq. \eqref{ps1} can be further approximated as

\beq\label{ps2}
\Delta^2 (k)  \simeq \frac{9}{8 \pi^2} \frac{k^4}{\rho_\phi (C  \tc)^4 } \left[ 1+ 2 \beta_k  \right]  \left[ \lambda_2^2 N(z_k) +  9  \lambda_1^2  M(z_k) \right],
\eeq
where we defined

\beq
\beta_k \equiv  \frac{ k^2 M_P^2}{\rho_\phi  (C  \tc)^2},
\eeq
with $M(z_k)$ and $N(z_k)$ as expressed in Eqs. \eqref{Map1} and \eqref{Nap1}. Therefore, the condition $k^2 \ll  \rho_\phi (C \tc /M_P)^2$ implies $\beta_k \ll 1$. 
    
As mentioned earlier, $z_k \ll 1$ must be satisfied in order to ensure that  the mode has a proper wavelength bigger than the Hubble radius  when the collapse is triggered. Therefore, one can perform a series expansion of the functions $N(z_k)$ and $M(z_k)$ for $z_k \ll 1$, this is,

\beq
M(z_k) \simeq \frac{1}{\sigma_k^4} \left( 1 - \frac{z_k^2}{3} \right) \qquad \textrm{and} \qquad N(z_k) \simeq \frac{1}{\sigma_k^4}  \frac{z_k^4}{9}.
\eeq

Now let us focus on the collapse scheme where the momentum variable collapse but not the field variable, i.e. the scheme were $\lambda_1=0$ and $\lambda_2 = 1$. In such case, the power spectrum takes the form

\barr
\Delta^2 (k) &\simeq & \frac{1}{8 \pi^2} \frac{1}{\rho_\phi (\eta_r C )^4 } \left[ 1+ 2 \beta_k  \right] {k^4},
\earr
where we used the definition $z_k \equiv k \tc$. The power spectrum is of the form $k^4$ and the dominant term does not contain any  parameter that can be adjusted to recover a nearly scale-independent spectrum. Thus, in the limit where $\beta_k \ll 1$ and $\lambda_1=0$ and $\lambda_2 = 1$ one cannot recover the standard prediction. 

Next, we focus on the scheme $\lambda_1 = 1$, $\lambda_2 = 0$. For this scheme 
    
\beq
\Delta^2 (k) \simeq \frac{9}{8 \pi^2} \frac{k^4}{\rho_\phi (C  \tc)^4 } \left[ 1+ 2 \beta_k  \right] \frac{9}{\sigma_k^4}  \left[ 1 - \frac{z_k^2}{3} \right]
\eeq

Substituting $\beta_k$ and $z_k$ in the last expression, the power spectrum is written explicitly  as

 \beq\label{psfinal}
 \Delta^2 (k) \simeq \frac{81}{8 \pi^2} \frac{1}{\rho_\phi (\eta_r C  \tc)^4 } \left[ 1 + k^2 \left( \frac{2M_P^2}{\rho_\phi C^2 {\tc}^2} - \frac{{\tc}^2}{3} \right)  \right].
 \eeq
 Hence, if $\tc$ is independent of $k$, i.e. the time of collapse does not depend on the mode $k$, one can recover a flat spectrum plus (small) first order  corrections of the form $k^2$.   
 
 The next step is to check if the amplitude of the spectrum [Eq. \eqref{psfinal}] is consistent with the latest CMB observations \cite{Planckcls13}. This is, the model  must satisfy that
 
 \beq\label{amplitudps}
 \frac{81}{8 \pi^2} \frac{1}{\rho_\phi (\eta_r C  \tc)^4 } \simeq 10^{-9}.
 \eeq
 Using the numerical values for $C$ and $\eta_r$ the last condition is re-expressed as
 
 \beq\label{condamp}
 \rho_\phi^{-1} \simeq 10^{-120} {\tc}^4.
 \eeq
Furthermore, the condition $\beta_k \ll 1$  written explicitly is

    \beq\label{condbeta}
     \frac{ k^2 M_P^2}{\rho_\phi  (C  \tc)^2} \ll 1.
    \eeq
  Using once again the numerical values for $C$ and $\eta_r$ and taking the greatest value of  the relevant values for $ k \simeq 10^{-1}$ Mpc$^{-1}$, the condition \eqref{condbeta}, together with the condition on the amplitude \eqref{condamp},   establishes an upper bound on the time of collapse, namely

  \beq\label{condtc}
  \tc \ll 10^{-2} \textrm{Mpc}.
  \eeq
This is, the time of collapse must be approximately much before the epoch of nucleosynthesis. Additionally, condition \eqref{condtc} is consistent with the condition $k \tc \ll 1$  for the modes of observational interest. One further consistency check is to ensure that $\rho_\phi \ll \rho_{\textrm{rad}} (\tc)$ given that $\rho_\phi$ must satisfy Eq. \eqref{condamp}, which assures that the power spectrum posses the correct amplitude. Therefore, from Friedmann's equation

\beq
\rho_{\text{rad}} = \frac{3 M_P^2 \mH_c^2}{a_c^2} \simeq \frac{3 M_P^2}{C^2 {\tc}^4} \simeq \frac{3 M_P^2 10^{-120} \rho_\phi}{C^2},
\eeq
where in the last equality we used Eq. \eqref{condamp}. Inserting the the numerical values for $C$ and $M_P$ yields

\beq
\rho_\phi \simeq 10^{-5}  \rho_{\text{rad}} .
\eeq
Thus, is consistent with the requirement that $\rho_{\text{rad}} \gg \rho_\phi$.

   For the scheme $\lambda_1 = \lambda_2 = 1$, the power spectrum can be approximated as
  
  \barr
\label{psfinal2}
  \Delta^2 (k) &\simeq& \frac{81}{8 \pi^2} \frac{k^4}{\rho_\phi (C  \tc)^4 } \left[ 1+ 2 \beta_k  \right] \frac{1}{\sigma_k^4}  \left[ 1 - \frac{z_k^2}{3} +  \frac{z_k^4}{81}\right]. \nonumber \\
  \earr
  Thus, the dominant term is of the same form as the scheme described by $\lambda_1=1$ and $\lambda_2=0$, therefore, the analysis proceeds in an identical fashion.

  Now let us analyze the case $k^2 \gg \rho_\phi (C \tc /M_P)^2$, which now implies $\beta_k \gg 1$. Therefore, the power spectrum in Eq. \eqref{ps1} can be approximated by
  
  \beq
  \Delta^2 (k) \simeq \frac{9 \rho_\phi}{8 \pi^2 M_P^4} \left( 1+\frac{2}{\beta_k}   \right) \left( \lambda_2^2 N(z_k) +  9  \lambda_1^2  M(z_k) \right). 
  \eeq

 We focus first on the collapse scheme $\lambda_1=1$ and $\lambda_2 =0$. In this case, upon using the series expansion Eq. \eqref{Map1}, one obtains
 
 \barr
 \Delta^2 (k) &\simeq& \frac{81 \rho_\phi}{8 \pi^2 M_P^4 \eta_r^4} \left( 1+\frac{2}{\beta_k}   \right)  k^{-4} \left( 1 - \frac{z_k^2}{3} \right).  \nonumber \\
 \earr
We see that the dominant term of the approximation is proportional to $k^{-4}$ and does not depend on the time of collapse, henceforth, one cannot recover the standard spectrum. 

The collapse scheme described by $\lambda_1=0$ and $\lambda_2 =1$ yields an approximated power spectrum expressed as

\beq
\Delta^2 (k) \simeq \frac{9 \rho_\phi}{8 \pi^2 M_P^4} \left( 1+\frac{2}{\beta_k}   \right) \frac{1}{\sigma_k^4} \frac{z_k^4}{9}.
\eeq
    Substituting $\beta_k$ and $z_k$ we have
    
    \beq
\Delta^2 (k) \simeq   \frac{ \rho_\phi}{8 \pi^2 M_P^4} \frac{{\tc}^4}{\eta_r^4}  \left( 1+\frac{2 \rho_\phi  (C  \tc)^2}{k^2 M_P^2}   \right). 
    \eeq
  Thus, in this scheme, if the time of collapse is independent of the mode $k$, the model predicts a scale-invariant spectrum plus corrections of the form $k^{-2}$.  Additionally, for this scheme, we must check if  the predicted amplitude is consistent with the latest CMB observations \cite{Planckcls13}:
  
  \beq
  \frac{ \rho_\phi}{8 \pi^2 M_P^4} \frac{{\tc}^4}{\eta_r^4}  \simeq  10^{-9}.
  \eeq
    Therefore, by inserting the numerical values the relation between the energy density and the time of collapse is
    
    \beq\label{cond2}
    \rho_\phi^{-1}  \simeq 10^{-129} {\tc}^4.
    \eeq
   The condition $\beta_k \gg 1$ is thus,
   
   \beq
   \frac{k^2M_P^2}{\rho_\phi(C \tc)^2} \gg 1.
   \eeq
    Using Eq. \eqref{cond2} and the numerical values of $C,\eta_r,M_P$ and the lowest value for the mode of interest $k \simeq 10^{-6}$ Mpc$^{-1}$, one obtains that the time of collapse must satisfy
    
    \beq
    \tc \gg 10^8 \textrm{Mpc},
    \eeq
   which is 6 orders of magnitude greater than the time of decoupling; consequently this scheme is also ruled out. 
   
   Finally, the approximated power spectrum for  the last scheme corresponding to $\lambda_1=\lambda_2=1$, is
   
   \barr
   \Delta^2 (k) &\simeq&   \frac{81 \rho_\phi}{8 \pi^2 M_P^4 \eta_r^4}  \left( 1+\frac{2}{\beta_k}   \right) \left( 1 - \frac{z_k^2}{3} + \frac{z_k^4}{81} \right)      k^{-4}. \nonumber \\
   \earr
As we see, the dominant term in the expansion is of the form $k^{-4}$ and therefore the scheme is discarded. 
    
We end this section by summarizing the main conditions under which the model can reproduce an nearly scale independent power spectrum. 
  
  The first condition is that the collapse scheme must be such that the field variable is affected by the collapse, i.e. $\bra \hat{y}_{\nk} (\tc) \ket \neq 0$; the momentum variable can or cannot be affected by the collapse. The second condition is that the \textbf{time of collapse must be independent of $k$}, i.e., $\tc=\eta_c$ the same for all modes and satisfy $\eta_c \ll 10^{-2}$ Mpc; this is a reasonable range for the time of collapse, since it should occur before the nucleosynthesis stage. If those conditions are met, then the power spectrum is explicitly
  
  \beq\label{psfinal3}
  \Delta^2 (k)  \simeq \mathcal{A} C(k), 
  \eeq
  where
  
  \beq
  \mathcal{A} \equiv \frac{81}{8 \pi^2} \frac{1}{\rho_\phi C^4 \eta_r^4 {\eta_c}^4  },   
  \eeq

  \beq\label{Ck}
  C(k) \equiv  \left(1 + 2 \beta_k\right)\left\{\frac{ \sin^2 (k\eta_c)}{(k\eta_c)^2} +  \frac{\lambda_2^2}{9}  \left[ \half + \frac{1}{2(k\eta_c)^2} + \cos(2k\eta_c) \left( \half - \frac{1}{2 (k\eta_c)^2}\right)  -\frac{\sin(2k\eta_c)}{k\eta_c}  \right]\right\},
  \eeq
  with $\lambda_2$ either 1 or 0 and $\rho_\phi$ to be adjusted by the amplitude. Therefore, apparently we have constructed a viable model for generating the primordial curvature perturbation. It is a viable model in the sense that our theoretical prediction Eq. \eqref{psfinal3} has a consistent amplitude and is almost independent of $k$.

  
%

Let us remark that the prediction from our model [Eq. \eqref{psfinal3}] is different from the standard one $\Delta^2 (k) =  A_s k^{n_s -1}$; in particular, the dependence on $k$ is not similar. In our model the dependence on $k$ is explicitly contained in the function $C(k)$ [see Eq. \eqref{Ck}], while in the standard case is given by $k^{n_s-1}$. This difference can be explained in part by noting that we have considered a perfect de Sitter space-time for the inflationary regime. On the other hand, we could have performed our calculations in a quasi-de Sitter Universe during inflation and that would have yielded a collapse power spectrum of the form $\Delta^2 (k) \simeq \mathcal{A} \widetilde C(k) k^{n_s-1}$, i.e., we would have obtained a power spectrum that would depend on $k$ in two ways: The first would be given by having introduced the collapse hypothesis, reflected in the function $\widetilde C(k)$, and the second one would have to do with the quasi-de Sitter background during inflation, hence the factor $k^{n_s-1}$. Nevertheless, the functional dependence on $k$, given by the collapse hypothesis, would have not been substantially different from the one obtained in this paper, this is, $\widetilde C(k) \simeq C(k)$. Therefore, by relying on pure de Sitter inflation, we have simplified our calculations but also we have  retained the dependence on $k$, within the power spectrum,  that has to do only with the collapse hypothesis, consequently, the predicted power spectrum, Eq. \eqref{psfinal3}  is not exactly scale-invariant even if pure de Sitter inflation was used for calculations.

%
%


In the next section, we will study the effects of the collapse during the radiation era on the CMB temperature and polarization fluctuation spectrum by considering only the approximate scale-invariant spectrum given by Eq. \eqref{psfinal3} that relies on  the assumption that the time of collapse is independent of $k$, i.e. $\tc = A$.

\section{Effects on the CMB fluctuation spectrum and comparison with observational data}\label{camb}

In order to analyze the effects of a collapse of the wave function of the inflaton field during the radiation era on the CMB fluctuations power spectrum, let us first define the fiducial model, which will be taken just as a reference to discuss the results we obtain for the collapse models. The fiducial model is a $\Lambda$CDM model with the following cosmological parameters: baryon density in units of the critical density, $\Omega_B h^2=0.02214$; dark matter density in units of the critical density, $\Omega_{CDM} h^2=0.1187$; Hubble constant in units of ${\rm Mpc^{-1} km \ \ s^{-1}}$, $H_0=67.8$; reionization optical depth, $\tau=0.092$; and the scalar spectral index, $n_s=0.9608$. These are the best-fit values presented by the \emph{Planck} collaboration \cite{Plancklike13} using the CMB temperature data released by \emph{Planck}, the CMB polarization data reported by WMAP \cite{wmap9cls}, CMB temperature data for high values of $l$ reported by ACT \cite{ACT13} and SPT \cite{SPT12} and Baryon Acoustic Oscillations \cite{BAO1,BAO2,BAO3,BAO4}. 

\begin{figure}
\begin{center}
\includegraphics[scale=0.31,angle=-90]{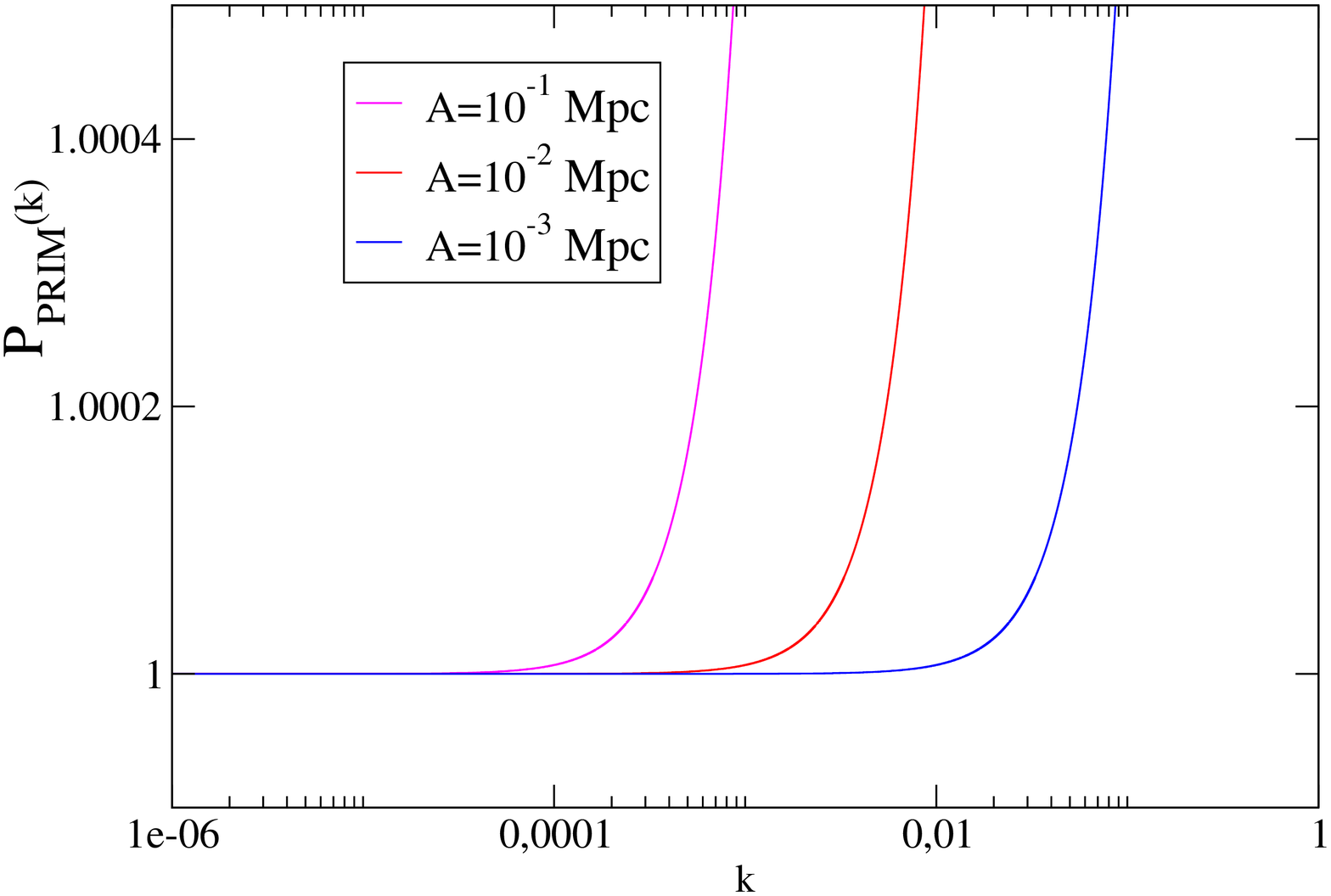}
\includegraphics[scale=0.31,angle=-90]{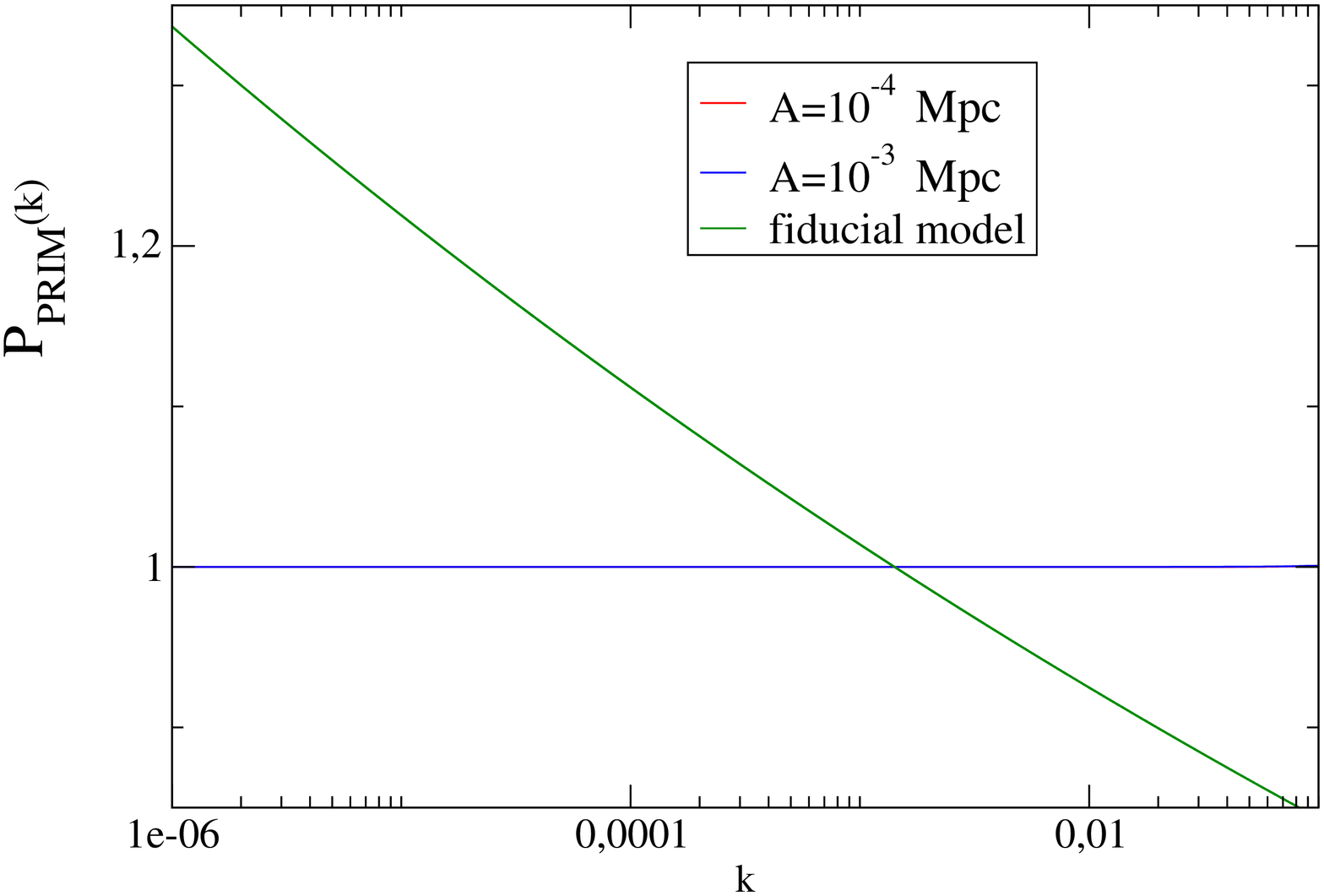}
\end{center}
\caption{Left: Primordial spectra, with wave function collapse of the inflaton field  during the radiation era, for different values of the collapse time $\tc=A$ and $\lambda_2=0$; Right: Primordial spectra with wave function collapse of the inflaton field during the radiation era ($\lambda_2=0$) and Primordial Spectra of the Fiducial Model (for these scales the collapse models are indistinguishable among themselves). }
\label{power}
\end{figure}

In Figure \ref{power} left, we show the primordial spectrum of models where a collapse of the wave function of the inflaton field during the radiation era has been included for different values of the collapse time $\tc= A $ and $\lambda_2=0$. It follows from Eq. \eqref{Ck} that the main contribution to $C(k)$ comes from the term $(1+2\beta_k) \simeq 1+ 10^5 z_k^2$ and thus setting $\lambda_2\neq 0$ does not change the primordial spectrum significantly. Therefore, we will only analyze the case $\lambda_2=0$ since the same conclusions apply to the case $\lambda_2\neq 0$. Fig. \ref{power} right, shows the primordial spectrum of the collapse models compared to the fiducial model. The variation between the collapse models due to different values of the collapse time is very tiny compared to the difference of these models with the fiducial model (see Fig. \ref{power} right). Thus, it follows  that the collapse models are very similar to a fiducial model with $n_s=1$ (which is ruled out at 5$\sigma$ by \emph{Planck}'s data)  and it will be difficult to fit these models to present data. 
This also reflects the fact, that if we would have considered quasi-de Sitter inflation, the shape of the collapse power spectrum during radiation and the one given by the standard single field slow-roll inflationary model, would have been, for all practical purposes, indistinguishable from each other. 
The main reason for this is the restriction $\tc \ll 10^{-2}$ Mpc that constrains the values of $A$ to be less than one and prevents the primordial power spectrum to  move over significantly from the standard power spectrum. This is not the case for the models where the collapse happens during inflation and therefore, we could find good fit to the WMAP data in previous works \cite{LSS12} and also to provide  features in the collapse power spectrum that made it distinguishable from the traditional spectrum. 
 

\begin{figure}
\begin{center}
\includegraphics[scale=0.5,angle=-90]{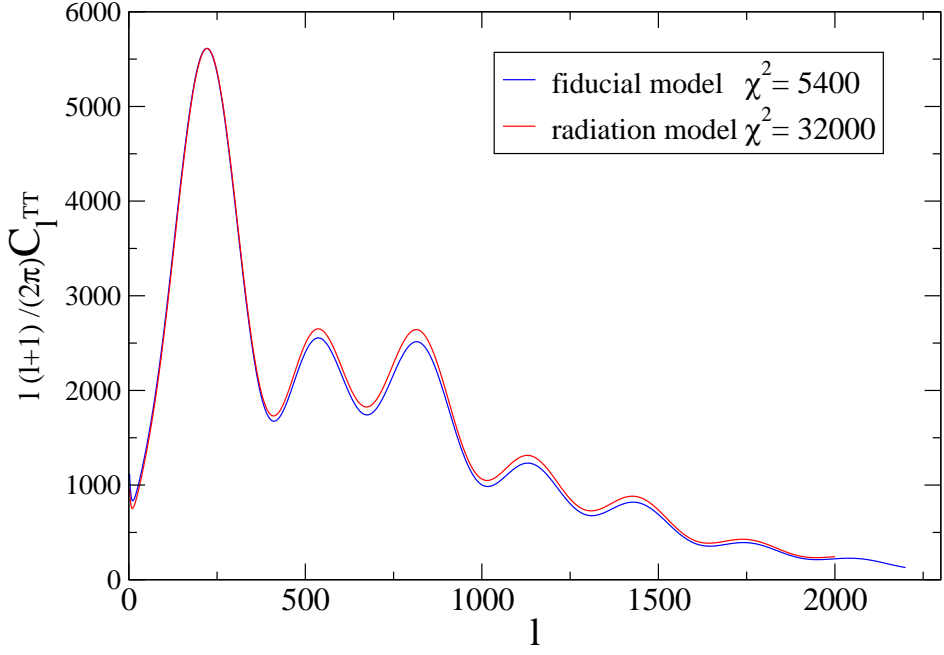}
\end{center}
\caption{The temperature auto-correlation (TT) power spectrum for  the fiducial model and for a model where the collapse of the inflaton wave function  happens during the radiation era at conformal time $\tc=10^{-3} \textrm{Mpc}$. All models are normalized  to the maximum of the first peak of the fiducial model. The value of  $\chi^2$ is calculated using WMAP9, \emph{Planck}, SPT and ACT release data (both temperature and temperature-polarization power spectrum are included).}
\label{clstt}
\end{figure}   

\begin{figure}
\begin{center}
\includegraphics[scale=0.31,angle=-90]{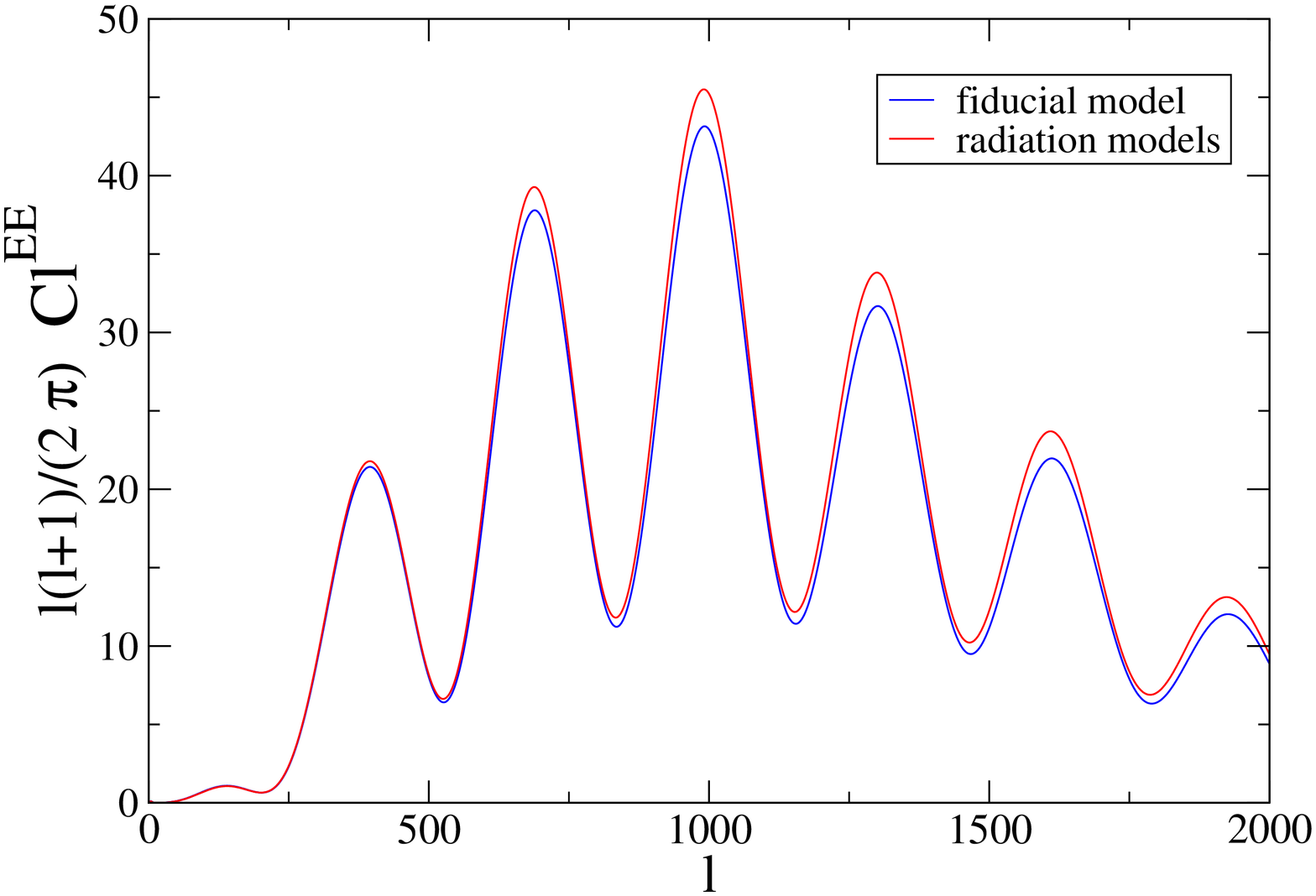}
\includegraphics[scale=0.31,angle=-90]{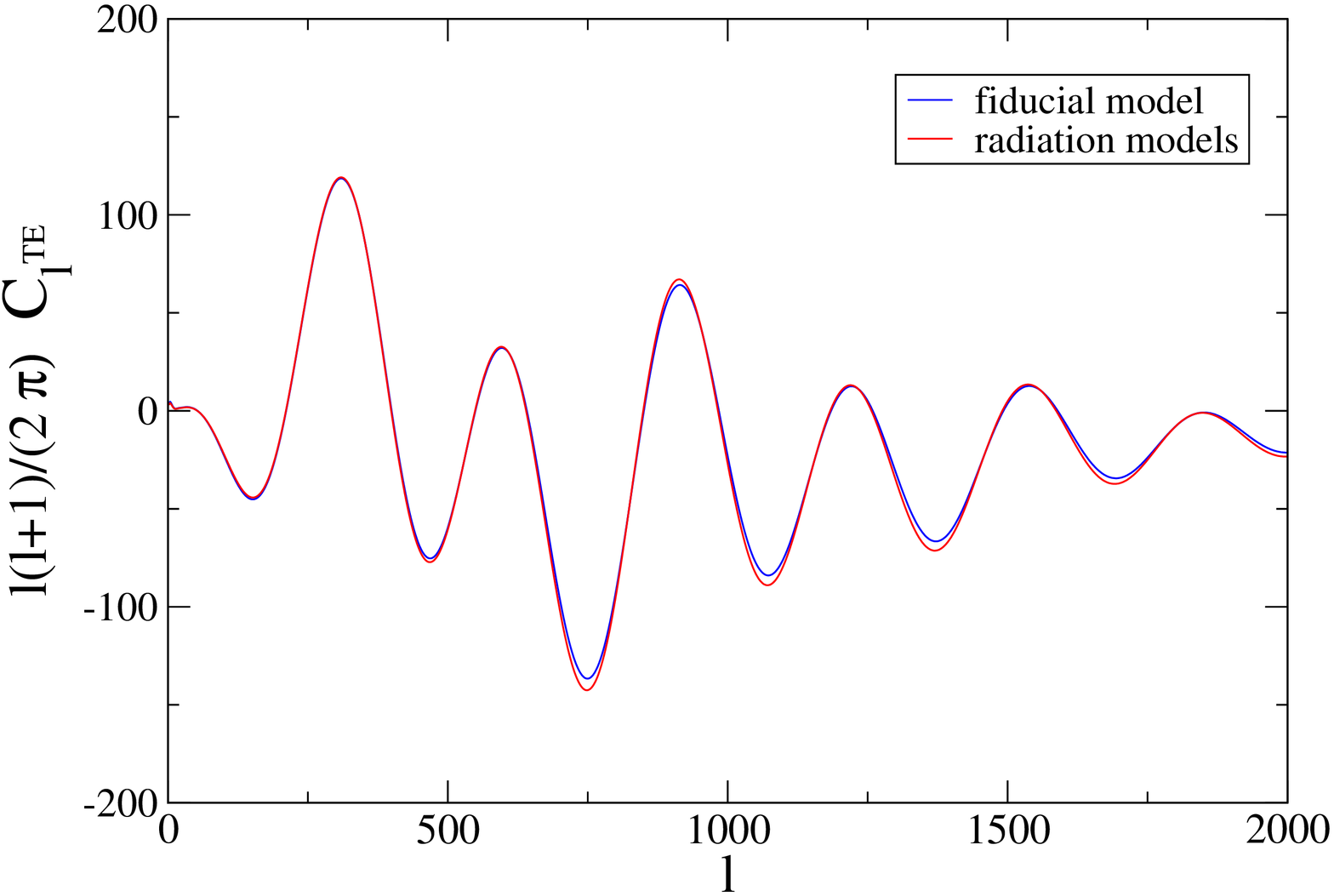}
\end{center}
\caption{Left: E polarization auto-correlation (EE) power spectrum; Right:  Temperature-polarization cross correlation (TE) power spectra. In both cases we plot the fiducial model and a model where the collapse of the inflaton wave function  happens during the radiation era at conformal time $\tc=10^{-3} \textrm{Mpc}$. All models are normalized  to the maximum of the first peak of the fiducial model. The value of  $\chi^2$ for all models is the same as indicated in Fig. \ref{clstt}.}
\label{clsEE}
\end{figure}   

Fig. \ref{clstt} shows the temperature auto-correlation power spectrum for the fiducial model and for the model where the collapse occurs during the radiation era. The respective EE and TE polarization power spectrum are shown in Fig. \ref{clsEE}. For all models satisfying the constraint $\tc \ll 10^{-2} \textrm{ Mpc}$, the temperature, the E polarization and the TE cross correlation power spectrum are the same as the one shown in Figs. \ref{clstt} and \ref{clsEE}, labeled as ``radiation models.'' The main reason for this, is the tiny difference in the primordial power spectrum for different radiation-collapse models shown in Fig. \ref{power}. The difference between the value of  $\chi^2$ for the fiducial and collapse models is significant ($\chi^2$ is calculated using WMAP9 polarization data, \emph{Planck} temperature data, SPT and ACT temperature data) and shows that a good fit to these data would be difficult to find for the collapse-radiation models. This is due to the low errors and accuracy of the present CMB data set. However, and in order to be sure about our conclusions, we intended to perform a statistical analysis to fit the CMB temperature power spectrum reported by the \emph{Planck} \cite{Planckcls13} collaboration and the polarization spectra reported by the WMAP \cite{wmap9cls} collaboration together with the temperature power spectrum for high $l$ from ACT \cite{ACT13} and SPT \cite{SPT12} and Baryon Acoustic Oscillations \cite{BAO1,BAO2,BAO3,BAO4}.  We performed our statistical analysis by exploring the parameter space with Monte Carlo Markov chains generated with the publicly available CosmoMC  code of Ref. \cite{LB02} that uses the Boltzmann code CAMB \cite{LCL00} to compute the CMB power spectra. We modified the primordial power spectrum according to Eq. \eqref{psfinal3} with $C(k)$ as given in Eq. \eqref{Ck} and with the time of collapse parameterized as $\tc = A$.  The parameters allowed to
vary are:
\begin{equation}
P=\left(\Omega_B h^2, \Omega_{CDM} h^2, \Theta, \tau, A_s, A\right),
\end{equation}
where $\Theta$ is the ratio of the comoving sound horizon at decoupling to the angular diameter distance to the surface of last scattering, $\tau$ is the reionization optical depth, $A_s$ is the amplitude of the primordial density fluctuations, and $A$ is the model's parameter related to the conformal time of collapse.  According to the previous discussion, we could not find a good convergence of the Markov chains, even more, the code got stuck about $200$ steps and/or  failed due to the value of the optical depth. This happens, because, in order to get a fit to the data, the code explores other values for the cosmological parameters far from the fiducial model.


Note that in Figs. \ref{power}, \ref{clstt} and \ref{clsEE}, the fiducial model assumed $n_s = 0.9608$, while for the collapse model we set $n_s =1$. If we would have considered a quasi-de Sitter inflation for our model instead of a pure de Sitter one, we should have set $n_s = 0.9608$ for our model too, but, as argued in the previous section, we could have still used the collapse power spectrum given by Eq. \eqref{psfinal3} since it should not be substantially different from the one obtained using quasi-de Sitter inflation. Therefore, as can bee seen in all figures, our model's prediction would have been practically the same as the fiducial one, which corresponds to the conventional inflationary scenario, both with $n_s = 0.9608$.

\section{Summary and Conclusions}\label{discussion}

In this paper we have constructed a plausible model for generating the primordial curvature perturbation during the radiation dominated era,  by assuming a self-induced collapse of the wave function associated to each mode of the inflaton field. In Section \ref{analisis}, we showed that there are two major conditions  for this model to be considered viable: i) the collapse must affect the perturbation of the inflaton field  while the respective momentum can or not be affected; ii) the time of collapse $\tc$ must be independent of the mode $k$. If these conditions are met, then our model predicts a nearly scale-invariant power spectrum, which in principle has a different shape from the one given by the conventional single-field slow-roll inflationary model. This difference in the shape of the power spectrum is exclusively provided by having introduced the collapse hypothesis and is reflected in the function $C(k)$ [see Eqs. \eqref{psfinal3}, \eqref{Ck}]. However, in Section \ref{camb} we showed that the changes to the primordial spectrum introduced by the collapse are very small. Moreover,  the  angular temperature and temperature-polarization CMB power spectrum, within the collapse proposal,  are essentially indistinguishable from the standard inflationary model in an exact de Sitter background. The fact that the angular power spectrum cannot be distinguished from the standard inflationary model arises from the requirement that the primordial power spectrum  matches the amplitude of scalar fluctuations consistent with the latest CMB observations. This latter requirement implies a constraint on the time of collapse $\tc \ll 10^{-2}$ Mpc. On the other hand, this constraint is consistent with the requisite that  the energy density of the inflaton field should be negligible compared with the energy density of the radiation field, if the collapse is supposed to take place in the radiation era. The restriction on the time of collapse, thus, does not allow the model's predictions  to depart too much from the standard ones. Additionally, considering a quasi-de Sitter background for the calculation of the inflaton perturbations during inflation, would have resulted in a primordial power spectrum equal to the fiducial model one with very small corrections due to the collapse of the inflaton's wave function. Therefore, the calculations performed in this paper, let us assure that the predictions of this model (using a quasi-de Sitter background for the calculations during inflation) for the CMB temperature and polarization fluctuation spectrum will not be different from the standard model ones. We would like to emphasize that this case is different from the one in which the collapse takes place during inflation and the changes in the primordial power spectrum due to the collapse hypothesis are important even in a perfect de Sitter background.  


\acknowledgments

The authors thank D. Sudarsky for useful discussions. Support for this work was provided by PIP 0152/10 CONICET. GL acknowledges financial support by CONICET postdoctoral grant. We also thank the referee for useful suggestions.

\bibliography{bibliografia}
\bibliographystyle{apsrev}

\end{document}